\documentclass{emulateapj}
\usepackage{natbib}
\usepackage{graphicx}
\usepackage{times}
\usepackage{xspace} 
\usepackage{booktabs,longtable,tabu}
\usepackage{tabularx}

\newcommand {\apgt} {\ {\raise-.5ex\hbox{$\buildrel>\over\sim$}}\ }
\newcommand {\aple} {\ {\raise-.5ex\hbox{$\buildrel<\over\sim$}}\ }

\newcommand {\chandra} {{\it Chandra}\xspace}
\newcommand {\xmm} {{\it XMM-Newton}\xspace}
\newcommand {\nustar} {{\it NuSTAR}\xspace}

\newcommand {\integral} {{\it INTEGRAL}\xspace}
\newcommand {\swiftxrt} {{\it Swift}~XRT\xspace}
\newcommand {\swiftbat} {{\it Swift}~BAT\xspace}

\newcommand {\wise} {{\it WISE}\xspace}

\newcommand {\sixum} {$6$~$\mu$m\xspace}
\newcommand {\Lsixum} {$L_{\rm 6\mu m}$\xspace}

\newcommand {\ewfeka} {$\mathrm{EW}_{\mathrm{FeK\alpha}}$\xspace}
\newcommand {\feka} {Fe~K$\alpha$\xspace}

\newcommand {\brnu} {$\mathrm{BR}_{\mathrm{Nu}}$\xspace}

\newcommand {\nh} {$N_{\mathrm{H}}$\xspace}

\newcommand {\lx} {$L_{\mathrm{X}}$\xspace}

\newcommand{\nii}{\mbox{[\ion{N}{2}]}\xspace}
\newcommand{\oi}{\mbox{[\ion{O}{1}]}\xspace}
\newcommand{\oii}{\mbox{[\ion{O}{2}]}\xspace}
\newcommand{\oiii}{\mbox{[\ion{O}{3}]}\xspace}

\newcommand{\ciii}{\mbox{\ion{C}{3}]}\xspace}
\newcommand{\civ}{\mbox{\ion{C}{4}}\xspace}

\newcommand{\sii}{\mbox{[\ion{S}{2}]}\xspace}

\newcommand{\halpha}{H$\mathrm{\alpha}$\xspace}
\newcommand{\hbeta}{H$\mathrm{\beta}$\xspace}

\newcommand{\typeii}{Type~2\xspace}

\newcommand {\ergpersec} {erg~s$^{-1}$\xspace}
\newcommand {\fluxunit} {erg~s$^{-1}$~cm$^{-2}$\xspace}
\newcommand {\fnuunit} {erg~s$^{-1}$~cm$^{-2}$~Hz$^{-1}$\xspace}
\newcommand {\nhunit} {cm$^{-2}$\xspace}
\newcommand {\degrees} {$^{\circ}$\xspace}

\newcommand {\wavdetect}{$\mathtt{wavdetect}$\xspace}
\newcommand {\xspec}{{\sc Xspec}\xspace}
\newcommand {\bntorus}{{\sc BNTorus}\xspace}

\newcommand {\pexrav}{$\mathtt{PEXRAV}$\xspace}

\newcommand {\transmission}{$\mathtt{transmission}$\xspace}
\newcommand {\reflection}{$\mathtt{reflection}$\xspace}
\newcommand {\torus}{$\mathtt{torus}$\xspace}

\begin{document}
%%%%%%%%%%%%%%%%%%%%%%%%%%%%%%%%%%%%%%%%%%%%%%%%%%%%%%%%%%%%%%%%%%%%%%

%%%%%%%%%%%%%%%%%%%%%%%%%%%%%%%%%%%%%%%%%%%%%%%%%%%%%%%%%%%%%%%%%%%%%%
\title{The {\it NuSTAR} Serendipitous Survey: Hunting for 
  The Most Extreme Obscured AGN at $>10$~keV}
%%%%%%%%%%%%%%%%%%%%%%%%%%%%%%%%%%%%%%%%%%%%%%%%%%%%%%%%%%%%%%%%%%%%%%

%%%%%%%%%%%%%%%%%%%%%%%%%%%%%%%%%%%%%%%%%%%%%%%%%%%%%%%%%%%%%%%%%%%%%%
\author{G.~B.~Lansbury\altaffilmark{1,2,$\dagger$}, 
D.~M.~Alexander\altaffilmark{2},
J.~Aird\altaffilmark{1},
P.~Gandhi\altaffilmark{3},
D.~Stern\altaffilmark{4},
M.~Koss\altaffilmark{5}, 
I.~Lamperti\altaffilmark{6},
M.~Ajello\altaffilmark{7},
A.~Annuar\altaffilmark{2},
R.~J.~Assef\altaffilmark{8},
D.~R.~Ballantyne\altaffilmark{9},
M.~Balokovi\'c\altaffilmark{10},
F.~E.~Bauer\altaffilmark{11,12,13},
W.~N.~Brandt\altaffilmark{14,15,16},
M.~Brightman\altaffilmark{10},
C.-T.~J.~Chen\altaffilmark{14},
F.~Civano\altaffilmark{17,18},
A.~Comastri\altaffilmark{19},
A.~Del~Moro\altaffilmark{20},
C.~Fuentes\altaffilmark{21},
F.~A.~Harrison\altaffilmark{10},
S.~Marchesi\altaffilmark{7},
A.~Masini\altaffilmark{19,22},
J.~R.~Mullaney\altaffilmark{23},
C.~Ricci\altaffilmark{11,24},
C.~Saez\altaffilmark{25},
J.~A.~Tomsick\altaffilmark{26},
E.~Treister\altaffilmark{21,11},
D.~J.~Walton\altaffilmark{1,4,10},
L.~Zappacosta\altaffilmark{27}
}
%%%%%%%%%%%%%%%%%%%%%%%%%%%%%%%%%%%%%%%%%%%%%%%%%%%%%%%%%%%%%%%%%%%%%%

%%%%%%%%%%%%%%%%%%%%%%%%%%%%%%%%%%%%%%%%%%%%%%%%%%%%%%%%%%%%%%%%%%%%%%
% Lansbury_1 & Aird:
\affil{$^{1}$Institute of Astronomy, University of Cambridge,
  Madingley Road, Cambridge, CB3 0HA, UK;
  $^{\dagger}$gbl23@ast.cam.ac.uk}
% Lansbury_1 & Alexander & Annuar:
\affil{$^{2}$Centre for Extragalactic Astronomy, Department of Physics,
  Durham University, South Road,
  Durham, DH1 3LE, UK}
% Gandhi:
\affil{$^{3}$School of Physics and Astronomy, University of
  Southampton, Highfield, Southampton SO17 1BJ, UK}
% Stern:
\affil{$^{4}$Jet Propulsion Laboratory, California Institute of
  Technology, 4800 Oak Grove Drive, Mail Stop 169-221, Pasadena, CA
  91109, USA}
% Koss:
\affil{$^{5}$Institute for Astronomy, Department of Physics, ETH
  Zurich, Wolfgang-Pauli-Strasse 27, CH-8093 Zurich, Switzerland}
% Lamperti:
\affil{$^{6}$Department of Physics and Astronomy, University College London, Gower Street, London, WC1E 6BT, UK}
% Ajello & Marchesi:
\affil{$^{7}$Department of Physics and Astronomy, Clemson University,
  Clemson, SC 29634-0978, USA}
% Assef:
\affil{$^{8}$N\'ucleo de Astronom\'ia de la Facultad de Ingenier\'ia,
  Universidad Diego Portales, Av. Ej\'ercito Libertador 441, Santiago,
  Chile}
% Ballantyne:
\affil{$^{9}$Center for Relativistic Astrophysics, School of Physics,
  Georgia Institute of Technology, Atlanta, GA 30332, USA}
% Balokovic & Harrison & Brightman:
\affil{$^{10}$Cahill Center for Astrophysics, 1216 East California
  Boulevard, California Institute of Technology, Pasadena, CA 91125,
  USA}
% Bauer_1 & Ricci_1 & Treister_2:
\affil{$^{11}$Instituto de Astrof{\'{\i}}sica and Centro de Astroingenier{\'{\i}}a, Facultad de F{\'{i}}sica, Pontificia Universidad Cat{\'{o}}lica de Chile, Casilla 306, Santiago 22, Chile}
% Bauer_2:
\affil{$^{12}$Millennium Institute of Astrophysics, Vicu\~{n}a Mackenna 4860, 7820436 Macul, Santiago, Chile}
% Bauer_3:
\affil{$^{13}$Space Science Institute, 4750 Walnut Street, Suite 205,
  Boulder, Colorado 80301, USA}
% Brandt_1 & Chen:
\affil{$^{14}$Department of Astronomy and Astrophysics, 
  The Pennsylvania State University, University Park, PA 16802, USA}
% Brandt_2:
\affil{$^{15}$Institute for Gravitation and the Cosmos, The Pennsylvania
State University, University Park, PA 16802, USA}
% Brandt_3:
\affil{$^{16}$Department of Physics, The Pennsylvania State University,
University Park, PA 16802, USA}
% Civano_1:
\affil{$^{17}$Yale Center for Astronomy and Astrophysics, Physics
  Department, Yale University, New Haven, CT 06520, USA}
% Civano_2:
\affil{$^{18}$Harvard-Smithsonian Center for Astrophysics, 60 Garden Street,
  Cambridge, MA 02138, USA}
% Comastri & Masini_1:
\affil{$^{19}$INAF-Osservatorio Astronomico di Bologna via Gobetti 93/3 40129, Bologna. Italy}
% Del Moro:
\affil{$^{20}$Max-Planck-Institut f\"ur Extraterrestrische Physik
  (MPE), Postfach 1312, D85741, Garching, Germany}
% Fuentes & Treister_1:
\affil{$^{21}$Universidad de Concepci\'{o}n, Departamento de
  Astronom\'{\i}a, Casilla 160-C, Concepci\'{o}n, Chile}
%
% Masini_2:
\affil{$^{22}$Dipartimento di Fisica e Astronomia (DIFA), Universit\`{à}
  di Bologna, viale Berti Pichat 6/2, 40127 Bologna, Italy}
% Mullaney:
\affil{$^{23}$Department of Physics and Astronomy, The University of
  Sheffield, Hounsfield Road, Sheffield, S3 7RH, UK}
% Ricci_2:
\affil{$^{24}$Kavli Institute for Astronomy and Astrophysics, Peking
  University, Beijing 100871, China}
% Saez:
\affil{$^{25}$Observatorio Astron\'omico Cerro Cal\'an, Departamento de
Astronom\'ia, Universidad de Chile, Casilla 36-D, Santiago, Chile}
% Tomsick:
\affil{$^{26}$Space Sciences Laboratory, 7 Gauss Way, University of
  California, Berkeley, CA 94720-7450, USA}
% Zappacosta:
\affil{$^{27}$INAF Osservatorio Astronomico di Roma, via Frascati 33,
  00040 Monte Porzio Catone (RM), Italy}
%%%%%%%%%%%%%%%%%%%%%%%%%%%%%%%%%%%%%%%%%%%%%%%%%%%%%%%%%%%%%%%%%%%%%%

%%%%%%%%%%%%%%%%%%%%%%%%%%%%%%%%%%%%%%%%%%%%%%%%%%%%%%%%%%%%%%%%%%%%%%

%%%%%%%%%%%%%%%%%%%%%%%%%%%%%%%%%%%%%%%%%%%%%%%%%%%%%%%%%%%%%%%%%%%%%%
\begin{abstract}

We identify sources with extremely hard X-ray spectra (i.e., with photon
indices of $\Gamma\lesssim 0.6$) in the $13$~deg$^{2}$ \nustar
serendipitous survey, to search for the most highly obscured AGNs
detected at $>10$~keV.
Eight extreme \nustar sources are identified, and we use the 
\nustar data in combination with lower energy X-ray observations (from \chandra, \swiftxrt,
and \xmm) to characterize the broad-band ($0.5$--$24$~keV) X-ray spectra. 
We find that all of the extreme sources are highly obscured AGNs, including three robust Compton-thick (CT; $N_{\rm H}> 1.5\times
10^{24}$~\nhunit) AGNs at low redshift ($z<0.1$), and a likely-CT AGN at higher redshift ($z= 0.16$). 
Most of the extreme sources would not have been identified as highly obscured
based on the low energy ($<10$~keV) X-ray coverage alone.
The multiwavelength properties (e.g., optical spectra and X-ray--MIR
luminosity ratios) provide further support for the eight sources
being significantly obscured. Correcting for absorption, the intrinsic
rest-frame $10$--$40$~keV luminosities of the extreme sources
cover a broad range, from $\approx 5\times 10^{42}$ to
$10^{45}$~\ergpersec.
The estimated number counts of CT AGNs in the \nustar serendipitous
survey are in broad agreement with model expectations based on previous X-ray surveys,
except for the lowest redshifts ($z<0.07$) where we measure a
  high CT fraction of $f_{\rm CT}^{\rm obs}=30^{+16}_{-12}\%$.
For the small sample of CT AGNs, we find a high fraction of galaxy
major mergers ($50\pm 33\%$) compared to control samples
of ``normal'' AGNs.

\end{abstract}
%%%%%%%%%%%%%%%%%%%%%%%%%%%%%%%%%%%%%%%%%%%%%%%%%%%%%%%%%%%%%%%%%%%%%%

%%%%%%%%%%%%%%%%%%%%%%%%%%%%%%%%%%%%%%%%%%%%%%%%%%%%%%%%%%%%%%%%%%%%%%
%%%%%%%%%%%%%%%%%%%%%%%%%%%%%%%%%%%%%%%%%%%%%%%%%%%%%%%%%%%%%%%%%%%%%%

%%%%%%%%%%%%%%%%%%%%%%%%%%%%%%%%%%%%%%%%%%%%%%%%%%%%%%%%%%%%%%%%%%%%%%
\section{Introduction}
\label{Introduction}
%%%%%%%%%%%%%%%%%%%%%%%%%%%%%%%%%%%%%%%%%%%%%%%%%%%%%%%%%%%%%%%%%%%%%%

The majority of cosmic supermassive
black hole growth has occured in an obscured phase (e.g., see
\citealt{Brandt15} for a review), during which gas and dust cover the
central active galactic nucleus (AGN). Historically, the
importance of highly obscured AGNs has been
inferred from the shape of the extragalactic cosmic X-ray background (CXB),
the high energy hump of which (peaking at $\approx 20$--$30$~keV)
requires significant populations of either highly obscured or
reflection-dominated systems (e.g., \citealt{Setti89,Comastri95,Gilli07,Treister09}). 
Large population studies have now quantified the relative abundance
of obscured and unobscured black hole growth phases (e.g., \citealt{Aird15a,Buchner15}).
A substantial fraction of the growth appears to occur
during the most obscured ``Compton-thick'' (``CT'' hereafter) phases, where the
absorbing column density exceeds the inverse of the Thomson scattering
cross-section ($N_{\rm H}\gtrsim 1.5\times 10^{24}$~\nhunit). 
However, the intrinsic absorption distribution of AGNs has proven difficult to
constrain, especially at the highly obscured to CT end, where AGNs are
particularly challenging to identify.

Besides completing a census,
identifying the most highly obscured AGNs is crucial to our
understanding of the environment of supermassive black hole growth.
The unified model of AGNs (e.g., \citealt{Antonucci93, Urry95, Netzer15}), which
largely succeeds at describing
AGNs in the local universe, posits that unobscured, obscured, and CT
systems have intrinsically similar nuclear structures but are simply
viewed from different inclination angles. In tension with this model
(at least in its simplest form) are observational results which find
possible evidence for high merger fractions in highly obscured AGN samples
(e.g., \citealt{Kocevski15,DelMoro16,Koss16a,Ricci17}).
Furthermore, observations of the
clustering of AGNs find that obscured
and unobscured AGNs may inhabit different large-scale environments (e.g.,
\citealt{Donoso14,DiPompeo14,DiPompeo16,Allevato11,Allevato14}; but
see also \citealt{Mendez16,Ballantyne17}). 
These results may suggest that high AGN obscuration can be linked to
specific phases in
the galaxy-AGN co-evolutionary sequence (e.g.,
\citealt{Sanders88,Hopkins08,Alexander12}), potentially associated with
periods of rapid black hole growth (e.g.,
\citealt{Draper10,Treister10a}). 

A challenge in answering these questions is that most wavelength
regimes are subject to strong biases against detecting highly
obscured AGNs, due to a combination of: (i) line-of-sight extinction and
(ii) dilution by light from other (e.g., stellar) processes. 
Selection methods exist which are relatively unhindered by (i), such
as mid-infrared (MIR) color selection (e.g., \citealt{Lacy04,Stern05,Daddi07,Fiore08,Stern12,Mateos12}) and optical
spectroscopic selection based on high ionization emission
lines (e.g., \citealt{Zakamska03,Reyes08}). However, these techniques both suffer from (ii),
especially at sub-quasar luminosities, and both still require
X-ray followup of the AGNs to provide accurate measurements of the
line-of-sight gas column densities (e.g., \citealt{Vignali06,Civano07,Alexander08,Vignali10,Jia13,LaMassa14,DelMoro16}). 
Hard ($>10$~keV) X-ray observations, on the other hand, have the advantage
of very little dilution from other processes, and are relatively unaffected by
line-of-sight obscuring material up to CT levels of absorption.

For high redshift AGNs ($z\gtrsim 2$) soft X-ray telescopes (e.g.,
\chandra and \xmm) sample the rest-frame hard X-ray band,
and are therefore effective in identifying the features of CT
absorption (e.g., \citealt{Comastri11,Brightman14}). In the lower-redshift
universe, however, hard X-ray telescopes become essential. Large
(e.g., all-sky) surveys with non-focusing hard X-ray
missions (e.g., \swiftbat and \integral) have been important for the
identification of highly obscured
AGNs in the very local universe ($z<0.05$; e.g.,
\citealt{Burlon11,Vasudevan13,Ricci15,Koss16a,Akylas16}). 
Now, with the first
focusing hard X-ray mission (\nustar; \citealt{Harrison13}) it is
possible to study source populations that are approximately two orders of
magnitude fainter, thus extending to lower luminosities and higher
redshifts. 
The largest extragalactic survey being undertaken with \nustar is the
serendipitous survey (\citealt{Alexander13,Lansbury17}), which has covered $\approx 13$~deg$^{2}$ and
detected $497$ sources, $276$ of which have spectroscopic
redshifts. The areal coverage and sample size are large compared to
the dedicated \nustar extragalactic blank-field surveys (e.g., in the ECDFS and
COSMOS fields; \citealt{Mullaney15,Civano15}), making the serendipitous
survey well suited to the discovery of rare populations such as
CT AGNs. The latter have proven elusive in the \nustar surveys thus
far, with only $1$--$2$ high-confidence CT AGNs being identified overall
(e.g., \citealt{Civano15}; Del~Moro et al.\ 2017, submitted; Zappacosta et al.\ 2017, submitted).

In this paper, we search for the most extreme hard X-ray sources in the $40$-month \nustar
serendipitous survey sample, and as a result reveal new robust CT AGNs. Firstly,
we select the objects with the
highest \nustar band ratios, implying very hard spectral shapes and
hence the likely presence of heavy absorption. Although band
ratios only give a crude estimate of absorption, they are nevertheless an
effective way to isolate the most extreme outliers (e.g., \citealt{Koss16a}). Secondly, we
perform a detailed analysis of the X-ray and multiwavelength
properties of these extreme objects, and discuss how their properties
compare to those of the general AGN population. The paper is structured as follows.
Section \ref{selection} describes the selection of the eight extreme
objects from the \nustar serendipitous survey sample. Section
\ref{data} details the data used and the soft X-ray counterparts. In Section \ref{xray} we
characterize the X-ray spectra of the sources (Section
\ref{xray_modelling}), and present the results for the X-ray spectral
properties (Section \ref{xray_results}). In Section \ref{indirect} we
investigate potential independent estimates of the source obscuration
properties through indirect techniques. Section \ref{optical} presents
the optical properties of the sample, including a summary of the
optical spectral properties (Section \ref{optical_spectra}) and 
host galaxy imaging, with a focus on the frequency of galaxy mergers 
(Section \ref{host_galaxies}). In Section \ref{prevalence} we discuss
the CT AGNs and their implications for 
the prevalence of CT absorption within the broader hard X-ray
selected AGN population. Finally, our main
results are summarized in Section \ref{summary}.
The cosmology adopted is ($\Omega_{M}$, $\Omega_{\rm \Lambda}$,
$h$)$=$($0.27$, $0.73$, $0.70$).
All uncertainties and limits are quoted at the $90\%$ confidence
level (CL), unless otherwise stated.

%%%%%%%%%%%%%%%%%%%%%%%%%%%%%%%%%%%%%%%%%%%%%%%%%%%%%%%%%%%%%%%%%%%%%%
\section{The Sample of Extreme, Candidate Highly Obscured AGNs from the {\it
    NuSTAR} Serendipitous Survey}
\label{selection}
%%%%%%%%%%%%%%%%%%%%%%%%%%%%%%%%%%%%%%%%%%%%%%%%%%%%%%%%%%%%%%%%%%%%%%

\begin{figure}
\centering
\includegraphics[width=0.47\textwidth]{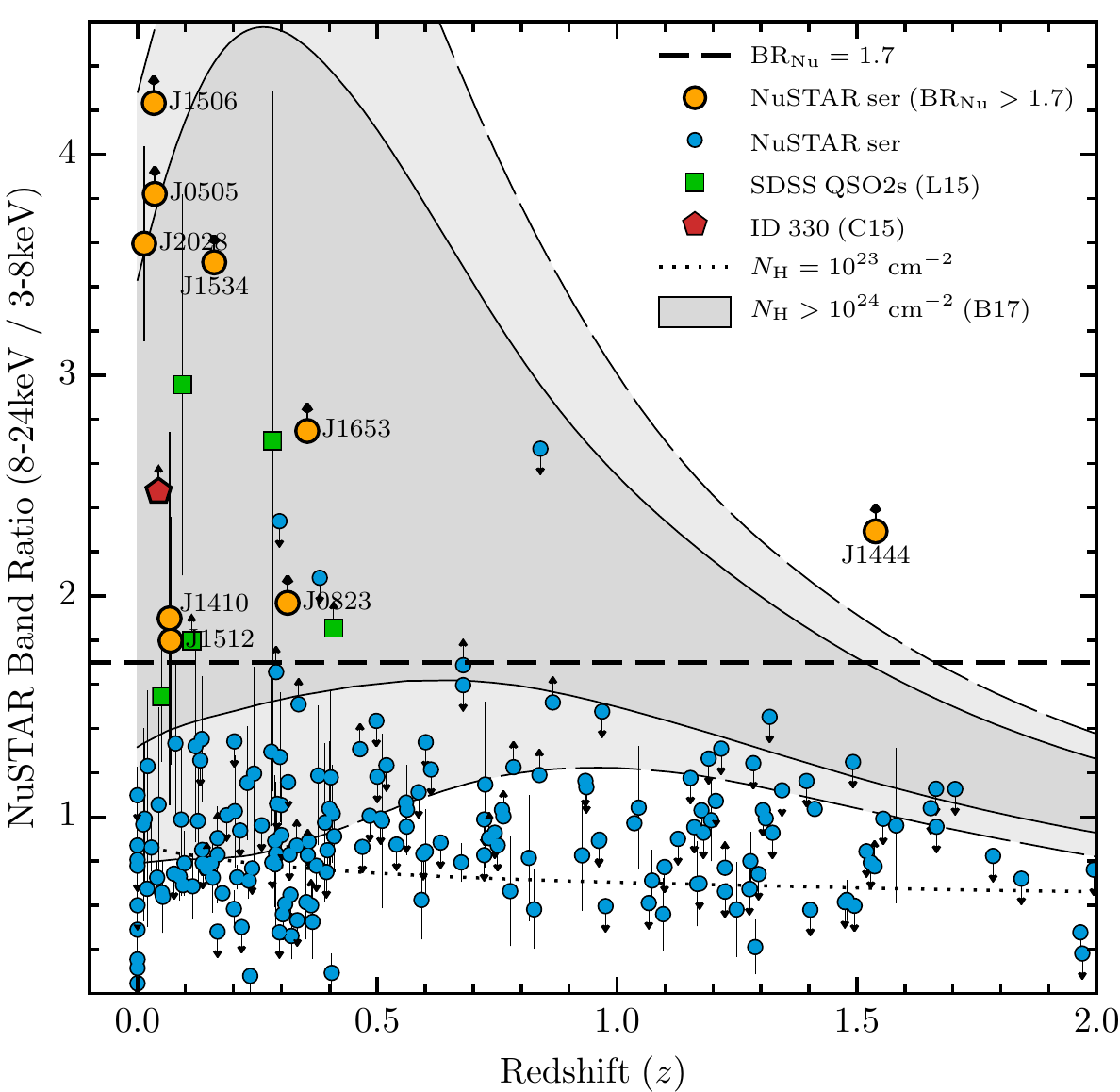}
\caption{\nustar band ratio (\brnu) as a function of redshift ($z$)
  for the \nustar serendipitous survey sample. The extremely hard
  ($\mathrm{BR_{Nu}}>1.7$; dashed line) serendipitous survey AGNs are
  shown as orange circles, and are individually labeled. ``Normal'' serendipitous survey
  sources at $\mathrm{BR_{Nu}}<1.7$ are shown as smaller blue circles.
  We compare to another extreme sample of optically (SDSS-) selected
  highly obscured Type~2 quasars observed with \nustar (green
  squares; \citealt{Lansbury14,Gandhi14,Lansbury15}), and to ID~330, the CT AGN
  identified in the \nustar-COSMOS survey (red pentagon;
  \citealt{Civano15}; Zappacosta et al.\ 2017, submitted). Additionally we compare to the expected band
  ratios for CT AGNs based on the high quality X-ray spectral modeling of very local CT AGNs in the
\nustar snapshot survey ($68\%$ percentiles in darker gray with solid borders; $90\%$
percentiles in lighter gray with long-dashed borders; \citealt{Balokovic14}; Balokovi\'{c} et al.\
2017, in prep.). 
For comparison, the dotted black curve shows the band ratios expected for a
moderate column density of $N_{\rm H}=10^{23}$~\nhunit.}
\label{br_z} 
\end{figure}

We start with the total $40$-month \nustar serendipitous survey sample
($497$ sources; \citealt{Lansbury17}).
To select sources with extremely hard X-ray spectra compared to the
rest of the \nustar serendipitous survey sample, we identify sources
with high hard-to-soft band ratios ($\mathrm{BR}_{\mathrm{Nu}}$),
calculated as the ratio of the $8$--$24$~keV to $3$--$8$~keV count rates.
We apply a cut at $\mathrm{BR}_{\mathrm{Nu}}> 1.7$ (see Figure
\ref{br_z}), which corresponds to an effective (i.e., observed) photon index of
$\Gamma_{\rm eff}\lesssim 0.6$.\footnote{The power law photon index ($\Gamma$)
  is defined as follows: $F_{E}\propto E^{-\Gamma}$, where $F_{E}$ is
  the photon flux and $E$ is the photon energy.} 
This cut is motivated by the
$\mathrm{BR}_{\mathrm{Nu}}$ values observed for
CT AGNs in other \nustar programs (e.g.,
\citealt{Balokovic14}; \citealt{Gandhi14}; \citealt{Civano15};
\citealt{Lansbury15}). 
We limit the sample to the sources with spectroscopic redshift
measurements, and exclude sources with upper limits in
$\mathrm{BR}_{\mathrm{Nu}}$.
The current spectroscopic completeness is $\approx 70\%$ for the
hard-band serendipitous survey sample at high galactic latitudes ($|b|
> 10$\degrees; \citealt{Lansbury17}).

Figure \ref{br_z} shows $\mathrm{BR}_{\mathrm{Nu}}$ versus redshift
for the \nustar serendipitous survey sample, excluding two sources
with erroneously high band ratios: NuSTARJ224225+2942.0, for which the
photometry is affected by contamination from a
nearby bright target; and NuSTARJ172805-1420.9, for which 
the photometry is unreliable due to a
high surface density of X-ray sources, with multiple \chandra sources
likely contributing to a blended \nustar detection (as determined using
\chandra data obtained through our followup program; PI J.\ A.\
Tomsick). 
Overall, nine sources have band ratios exceeding the selection
threshold of $\mathrm{BR_{Nu}}>1.7$ (all individually labelled in
Figure \ref{br_z}). 
We exclude NuSTAR~J202828+2543.4 (hereafter J2028; $z=0.01447$) 
from 
this work, as the source is closely
associated with the science target of the \nustar field
(IGRJ20286+2544; projected separation of $26$~kpc), and the
extreme obscuration and merger properties of this system are the focus of a detailed
study in \citet{Koss16b}. 
The main sample of eight extreme \nustar sources is listed in Table
\ref{nustar_basic_table}.

\renewcommand*{\arraystretch}{1.1}
\begin{table*}
\centering
\caption{The extremely hard \nustar serendipitous survey sources}
\begin{tabular}{lcccccccc} \hline\hline \noalign{\smallskip}
Full object name & Short name & R.A. & Decl. & $z$ & \brnu & Det. &
 $N_{\rm H,Gal}$ & Field name \\
(1) & (2) & (3) & (4)  & (5)  & (6) & (7) & (8) & (9) \\
\noalign{\smallskip} \hline \noalign{\smallskip}
NuSTAR\,J050559-2349.9 & J0505 & $76.49839$ & $-23.83169$ & $0.036$ & $>3.8$ & F H & $0.2$ & 2MASX\,J05054575-235113 \\
NuSTAR\,J082303-0502.7 & J0823 & $125.76385$ & $-5.04650$ & $0.313$ & $>2.0$ & F H & $0.5$ & FAIRALL\,0272 \\
NuSTAR\,J141056-4230.0 & J1410 & $212.73727$ & $-42.50139$ & $0.067$ & $1.9$$\pm 0.8$ & F S H & $0.5$ & 2MASX\,J14104482-422832 \\
NuSTAR\,J144406+2506.3 & J1444 & $221.02820$ & $25.10515$ & $1.539$ & $>2.3$ & F H & $0.3$ & PKS\,1441+25 \\
NuSTAR\,J150645+0346.2 & J1506 & $226.69040$ & $3.77118$ & $0.034$ & $>4.2$ & F H & $0.4$ & 2MASX\,J15064412+035144 \\
NuSTAR\,J151253-8124.3 & J1512 & $228.22497$ & $-81.40501$ & $0.069$ & $1.8$$\pm 0.6$ & F S H & $1.0$ & 2MASX\,J15144217-812337 \\
NuSTAR\,J153445+2331.5 & J1534 & $233.68763$ & $23.52593$ & $0.160$ & $>3.5$ & H & $0.4$ & Arp\,220 \\
NuSTAR\,J165346+3953.7 & J1653 & $253.44313$ & $39.89639$ & $0.354$ & $>2.7$ & H & $0.2$ & Mkn\,501 \\
\noalign{\smallskip} \hline \noalign{\smallskip}
\end{tabular}
\begin{minipage}[c]{0.97\textwidth}
\footnotesize
\textbf{Notes.} The sources are listed in order of increasing
right ascension. The entries in this table are drawn from the \nustar
serendipitous survey source catalog (\citealt{Lansbury17}). (1):
\nustar serendipitous source name. (2): Abbreviated \nustar source name adopted
in this paper. (3) and (4): Right ascension and
declination J2000 coordinates in decimal degrees. (5): Source spectroscopic
redshift. All redshifts are robust, except for J1444 where
  fewer lines are identified (see Section \ref{optical}). (6): \nustar photometric band ratio, as defined in Section \ref{selection}. (7): The \nustar energy bands for which the
source is independently detected. F, S, and H correspond to the full
($3$--$24$~keV), soft ($3$--$8$~keV), and hard ($8$--$24$~keV) bands,
respectively. 
(8): Line-of-sight Galactic column density
  (\citealt{Kalberla05}). Units: $10^{21}$~\nhunit.
(9): Name of the \nustar science target, in the field of which the
serendipitous source is detected.
\end{minipage}
\label{nustar_basic_table}
\end{table*}

Here we comment on the maximum energies at which the sources are
detected with \nustar.
Table \ref{nustar_basic_table} lists the standard \nustar energy
bands (i.e., the full, soft, and hard bands) for which sources are detected. 
By selection, all eight sources are detected in the $8$--$24$~keV band.
Splitting this hard band into sub-bands of $8$--$16$~keV and $16$--$24$~keV, all
eight sources are detected in the former band, and all except J1444 and J1653 are
detected in the latter band. For the six sources detected at
$16$--$24$~keV, the highest and lowest Poisson false probabilities are
$P_{\rm False}=2\times 10^{-3}$ and $10^{-8}$, respectively (i.e., the
detections range from $\approx 3\sigma$ to highly significant).
Only one source shows evidence for emission at
$>24$~keV: J1506, which is detected in the $24$--$50$~keV band at the
$\approx 3\sigma$ significance level. In summary, two sources are
detected up to a maximum energy of $\approx 16$~keV, five sources are detected up to $\approx 24$~keV,
and a single source is weakly detected at even higher energies (up to $\approx 50$~keV).

\subsection{A note on associated sources}
\label{associated_note}

Six out of eight ($75\%$) of the sources in this sample were 
serendipitously detected in \nustar observations of bright low-redshift \swiftbat AGNs.
The three serendipitous \nustar sources J0505, J1506, and J1512 are likely to be
weakly associated with the brighter BAT AGNs: each source lies
within $\pm 500~\mathrm{km/s}$ of the
redshift of the BAT AGN and at a projected separation of
$\approx 150$--$550$~kpc. 
The associations are ``weak'' in that the physical separations are
large, and the sources are therefore unlikely to be interacting.
The associated redshifts are unlikely to occur by chance given that
hard X-ray sources at these flux levels ($f_{\mathrm{8\mbox{-}24keV}}=
2$--$6\times 10^{-13}$~\fluxunit), and within $\pm 500~\mathrm{km/s}$ of
the BAT redshifts, have very low sky densities of $\approx
0.01$~$\mathrm{deg}^{-2}$ (e.g., \citealt{Treister09}). 
The latter implies low chance coincidence rates of $\approx
10^{-3.5}$ for the three cases of associated redshifts above. 
The effect of these weak associations on number counts measurements
for CT AGNs is accounted for in Section \ref{prevalence}.

In the overall $40$-month \nustar serendipitous survey,
  redshift associations between serendipitous sources and science targets like
  the above are rare (\citealt{Lansbury17}).\footnote{Sources are
    classed as associated if their velocity offset from
    the science target [$\Delta (cz)$] is smaller than $5\%$ of
    the total science target velocity (see \citealt{Lansbury17}).}
 The exception is at $z<0.07$ where, out of $15$
  sources in total, $10$ sources (including J0505, J1506, and J1512) 
  show evidence for associations.
We emphasise however that the majority of extragalactic sources in the \nustar
serendipitous survey ($247/262$ of the spectroscopically identified sample) are at
higher redshifts ($z>0.07$),\footnote{At $z>0.07$
only two sources are flagged as associated.} 
meaning that number counts measurements for the
survey (e.g., \citealt{Harrison16}) are not impacted.

%%%%%%%%%%%%%%%%%%%%%%%%%%%%%%%%%%%%%%%%%%%%%%%%%%%%%%%%%%%%%%%%%%%%%%
\section{Data}
\label{data}
%%%%%%%%%%%%%%%%%%%%%%%%%%%%%%%%%%%%%%%%%%%%%%%%%%%%%%%%%%%%%%%%%%%%%%

Table \ref{xrayData_table} provides details of the \nustar and soft 
($< 10$~keV) X-ray (i.e., \chandra, \swiftxrt, and \xmm) datasets
used in this work. 
For each source we
adopt the soft X-ray observatory data which provides the most sensitive
coverage at $<10$~keV. 
For four sources this coverage is from \swiftxrt, for three sources it is from
\xmm, and for one source it is from \chandra.
For five sources we use the combined soft X-ray dataset
from multiple individual observations (as detailed in Table
\ref{xrayData_table}) to obtain the most
precise X-ray constraints possible. The soft X-ray observations are
generally not contemporaneous with the \nustar observations. 
Section \ref{xray_modelling} discusses the possibility of variability for these
sources.

\renewcommand*{\arraystretch}{1.7}
\begin{table*}
\centering
\caption{Summary of the X-ray data adopted for the spectroscopic and
  photometric X-ray analyses}
\begin{tabular}{llllllllllll} 
\hline\hline \noalign{\smallskip}
\multicolumn{1}{l}{} & \multicolumn{5}{c}{\nustar Observations} &
\multicolumn{6}{c}{Soft X-ray Observations} \\
% \noalign{\smallskip}
\cmidrule(rl){2-6} \cmidrule(rl){7-12} 
% \noalign{\smallskip}
Object & Observation ID & UT Date & $t$ & $S_{\rm net}$ & $B$ &
                                                                Observatory & Observation ID & UT Date & $t$ & $S_{\rm net}$ & $B$ \\
(1) & (2) & (3) & (4)  & (5)  & (6) & (7) & (8) & (9) & (10)  & (11)  & (12) \\
\noalign{\smallskip} \hline \noalign{\smallskip}
J0505 & 60061056002 & 2013-08-21 & $21.1$ & $66$ & $53$ & {\it XMM-Newton} & \parbox[t]{1.5cm}{0605090101$^{c}$} & \parbox[t]{1.5cm}{2009-08-06} & \parbox[t]{0.4cm}{29.4} & \parbox[t]{0.4cm}{70} & \parbox[t]{0.4cm}{46} \\
J0823 & 60061080002$^{a}$ & 2014-01-10 & $24.3$ & $41$ & $67$ & {\it XMM-Newton} & \parbox[t]{1.5cm}{0501210501} & \parbox[t]{1.5cm}{2007-10-14} & \parbox[t]{0.4cm}{8.4} & \parbox[t]{0.4cm}{12} & \parbox[t]{0.4cm}{9} \\
J1410 & 60160571002 & 2015-05-14 & $22.2$ & $153$ & $125$ & {\it Swift} XRT & \parbox[t]{1.5cm}{00040973002 00040973003 00081157002 00081157003} & \parbox[t]{1.5cm}{2010-09-27 2011-03-10 2015-04-30 2015-05-14} & \parbox[t]{0.4cm}{5.0 5.0 5.8 5.6} & \parbox[t]{0.4cm}{$\cdots$ $\cdots$ $\cdots$ $\cdots$} & \parbox[t]{0.4cm}{$\cdots$ $\cdots$ $\cdots$ $\cdots$} \\
J1444 & 90101004002 & 2015-04-25 & $38.2$ & $62$ & $153$ & {\it Swift} XRT & \parbox[t]{1.5cm}{$\cdots$ 00033768001 00033768002 00033768003 00033768004 00033768005 00033768006} & \parbox[t]{1.5cm}{$\cdots\qquad$ 2015-05-13 2015-05-18 2015-06-01 2015-09-04 2016-04-13 2016-04-17} & \parbox[t]{0.4cm}{19.6$^{d}$ 3.1 3.0 4.1 4.0 4.0 1.4} & \parbox[t]{0.4cm}{10 $\cdots$ $\cdots$ $\cdots$ $\cdots$ $\cdots$ $\cdots$} & \parbox[t]{0.4cm}{$\cdots$ $\cdots$ $\cdots$ $\cdots$ $\cdots$ $\cdots$ $\cdots$} \\
J1506 & 60061261002 & 2014-09-08 & $21.3$ & $81$ & $122$ & {\it Swift} XRT & \parbox[t]{1.5cm}{00036622001 00036622002 00080144001} & \parbox[t]{1.5cm}{2007-12-19 2007-12-21 2014-09-08} & \parbox[t]{0.4cm}{9.4 8.7 6.1} & \parbox[t]{0.4cm}{$\cdots$ $\cdots$ $\cdots$} & \parbox[t]{0.4cm}{$\cdots$ $\cdots$ $\cdots$} \\
J1512 & 60061263002 & 2013-08-06 & $13.3$ & $153$ & $74$ & {\it Swift} XRT & \parbox[t]{1.5cm}{00036623001 00036623002 00080146001} & \parbox[t]{1.5cm}{2007-06-07 2007-06-09 2013-08-06} & \parbox[t]{0.4cm}{6.2 5.3 6.8} & \parbox[t]{0.4cm}{11 7 11} & \parbox[t]{0.4cm}{$\cdots$ $\cdots$ $\cdots$} \\
J1534 & 60002026002$^{b}$ & 2013-08-13 & $66.7$ & $42$ & $133$ & {\it Chandra} & \parbox[t]{1.5cm}{16092} & \parbox[t]{1.5cm}{2014-04-30} & \parbox[t]{0.4cm}{171.5} & \parbox[t]{0.4cm}{10} & \parbox[t]{0.4cm}{10} \\
J1653 & 60002024002$^{b}$ & 2013-04-13 & $18.3$ & $14$ & $16$ & {\it XMM-Newton} & \parbox[t]{1.5cm}{0652570101$^{c}$ 0652570201$^{c}$} & \parbox[t]{1.5cm}{2010-09-08 2010-09-10} & \parbox[t]{0.4cm}{43.7 44.0} & \parbox[t]{0.4cm}{73 82} & \parbox[t]{0.4cm}{47 42} \\
\noalign{\smallskip} \hline \noalign{\smallskip}
\end{tabular}
\begin{minipage}[c]{0.97\textwidth}
\footnotesize
\textbf{Notes.} (1): The abbreviated \nustar source name. (2) and (3):
The \nustar observation ID and start date (YYYY-MM-DD). (4), (5), and
(6): The net exposure
time (ks), net source counts, and scaled background counts, respectively,
for the extracted $3$--$24$~keV (or $8$--$24$~keV for J1534 and J1653)
\nustar spectrum. (7): The soft X-ray
observatory with the best (or in some cases, the only) coverage, which
we adopt for the analyses. 
(8) and (9): the adopted soft X-ray observation ID(s) and their
corresponding start date(s) (YYYY-MM-DD), respectively. (10), (11),
and (12): The exposure time (ks), net source counts, and scaled background
counts, respectively. For J0505, J0823, J1444, J1512, J1534, and J1653, these columns correspond to the extracted X-ray 
spectra (at $0.5$--$10$~keV, $0.6$--$10$~keV, and $0.5$--$8$~keV for
\xmm, \swiftxrt, and \chandra,
respectively). For the remaining two sources which are undetected at
soft X-ray energies (1410 and J1506), the \swiftxrt data tabulated
here are used for photometric constraints. 
$^{a}$: Here we use the \nustar FPMB data only (i.e., 
excluding the FPMA data).
$^{b}$: In these cases we limit the \nustar spectral analysis to
the $8$--$24$~keV band, since the sources are undetected in the soft
($3$--$8$~keV) and full ($3$--$24$~keV) \nustar bands, indicating no
significant source emission at $<8$~keV. 
$^{c}$: In these cases we use the combined MOS1+MOS2 data
only. $^{d}$: Here we quote the total exposure time and counts
(summing across all observations), since the source is
undetected in individual \swiftxrt observations.
\end{minipage}
\label{xrayData_table}
\end{table*}

\subsection{Soft X-ray counterparts to the extreme \nustar sources}
\label{xray_cparts}

\begin{figure*}
\centering
\includegraphics[width=\textwidth]{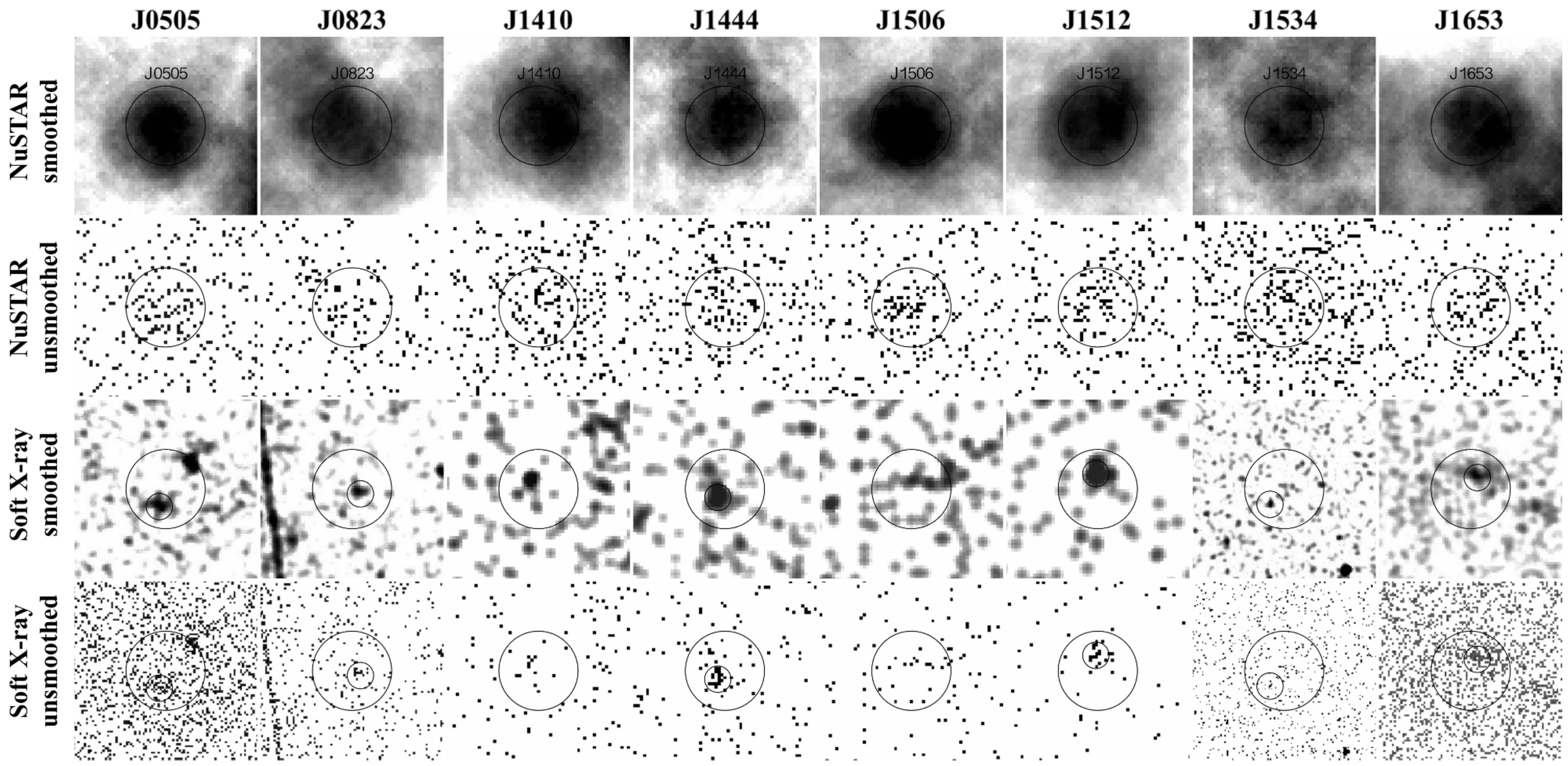}
\caption{\nustar and soft X-ray (\chandra, \swiftxrt, and \xmm) images
  for the eight extreme \nustar serendipitous
  survey sources. Each column corresponds to an individual \nustar source (the
  abbreviated source names are shown). 
$30''$-radius circular apertures are shown for
  each source, centered on the \nustar position.
  Upper two rows: \nustar hard ($8$--$24$~keV) band images, both
  smoothed (with a top hat function of radius $14$~pixels; first row) and
  unsmoothed (second row). Lower two rows: soft X-ray images from
  \chandra (the $0.5$--$2$~keV band is shown for J1534), \xmm (the
  full energy band is shown for J0505, J0823, and J1653), and
  \swiftxrt (the full energy band is shown for J1410, J1444, J1506,
  and J1512). The data are shown both smoothed (with a Gaussian function of radius
  $3$~pixels; third row) and unsmoothed (fourth row). The soft
  X-ray counterpart positions are marked by smaller ($10''$ radius)
  circular apertures, for all of the sources except J1410 and J1506
  (which are undetected in the available \swiftxrt coverage; see
  Section \ref{xray}). 
}
\label{ds9_xray} 
\end{figure*}

The soft X-ray counterparts improve the X-ray positional accuracy and,
when combined with the \nustar data,
allow for accurate spectral constraints using the broadest energy band
possible. Of the eight extreme \nustar sources studied here, two lack
soft X-ray counterparts (J1410 and J1506). In these cases there is no \chandra or \xmm coverage, and the
sources are undetected in the combined archival \swiftxrt
coverage (running \wavdetect with a detection threshold of $10^{-4}$).
The other six extreme \nustar sources have identified soft X-ray
counterparts. For five of these (J0505, J0823, J1444, J1512, and J1653) 
the soft X-ray counterparts are identified in \citet{Lansbury17}. Since J0505 has two \xmm sources nearby to the \nustar
source, we provide evidence below to support our correct counterpart
identification in this case.
For the remaining source (J1534), the \chandra counterpart is faint and
did not satisfy the detection criteria in \citet{Lansbury17}. Below we detail the identification of this specific
counterpart.

For J0505, there are two potential counterparts in the 3XMM
catalog, one at $14''$ offset from the \nustar position
($\mathrm{R.A.}=76.49983$\degrees, $\mathrm{decl.}=-23.83536$\degrees;
hereafter ``XMM1'') and one brighter source at $27''$ offset
($\mathrm{R.A.}=76.49296$\degrees $\mathrm{decl.}=-23.82597$\degrees;
hereafter ``XMM2''). To examine the X-ray spectra, we use the
MOS data for XMM1 (the source lies on a chip gap for PN)
and the PN plus MOS data for XMM2. The $0.5$--$10$~keV spectrum for XMM1 is
extremely flat (with an effective photon index of $\Gamma_{\rm
  eff}=-0.9^{+0.8}_{-1.4}$) and there is a
line detection consistent with \feka (rest-frame $E=6.3\pm0.1$~keV). The \feka
line has a high equivalent width of $\mathrm{EW}_{\rm Fe
  K\alpha}=1.4^{+1.4}_{-0.9}$~keV, suggesting a highly absorbed AGN. For XMM2, the $0.5$--$10$~keV spectrum is
steeper ($\Gamma_{\rm eff}=1.4\pm 0.2$). Although, XMM2 is brighter
than XMM1 over the full energy band, XMM1 is significantly brighter
for the energies at which \nustar is sensitive: for the $3$--$10$~keV
energy band, XMM1 and XMM2 have fluxes of $8.9\times 10^{-14}$~\fluxunit and
$1.8\times 10^{-14}$~\fluxunit, respectively. Given these fluxes and the relative spectral
slopes of XMM1 and XMM2 (with the former sharply increasing, and the
latter decreasing, towards higher X-ray energies), and the fact that the majority of \nustar
source counts ($79\%$) lie at high energies ($>8$~keV), we expect XMM1
to dominate the \nustar detected emission. We therefore adopt XMM1 as the
counterpart to J0505. 

For J1534, the deepest soft X-ray coverage is from a
 $171.5$~ks \chandra observation (obsID 16092, which targeted
 Arp~220). Running \wavdetect for the broad \chandra
 energy band of $0.5$--$7$~keV, no sources are blindly detected 
 within the \nustar error circle with
 false-probabilities (i.e., $\mathtt{sigthresh}$ values) of $P_{\rm
   False}\le 10^{-4}$. However,
 running the source detection for multiple energy bands, there is a
 significant detection at $0.5$--$2$~keV, with $P_{\rm
   False}\approx 10^{-6}$. 
Adding further confidence to the reliability of this source, SDSS coverage reveals a prominent $z = 0.160$ galaxy within the \nustar
error circle (SDSS~ J153445.80+233121.2), which agrees with
the \chandra position within the positional uncertainties ($0.6''$ offset). 
For an independent assessment of the significance of the \chandra
source, we perform aperture photometry ($2\arcsec$ source radius; large
background annulus) at the SDSS position. For the $0.5$--$2$~keV
band, the source is indeed detected at the $4.0\sigma$ level
(according to the binomial false probability). 
The \nustar/\chandra flux ratio for J1534 is extremely high
  (e.g., $f_{\rm  8\mbox{-}24}/f_{\rm 0.5\mbox{-}2}=141$). 
For comparison, four sources in the \nustar-COSMOS survey have
similarly high flux ratios (ranging from $f_{\rm  8\mbox{-}24}/f_{\rm
  0.5\mbox{-}2}=100$ to $304$), and all of these have been identified as
highly obscured AGNs (e.g., \citealt{Brightman14,Lanzuisi15a}; Zappacosta et
al., submitted).
The \chandra spectrum for J1534 is further discussed in Section \ref{xray_modelling}.

\subsection{X-ray spectroscopic products}

The NuSTARDAS task $\mathtt{nuproducts}$ is used to extract \nustar
source spectra, background spectra, and response files.\footnote{http://heasarc.gsfc.nasa.gov/docs/nustar/analysis} We adopt circular source extraction regions of
$45''$ radius where possible, and of $30''$ radius for two cases where the source is
either close to a bright science target or to the FoV edge. 
We perform separate spectral extractions for the two individual
\nustar telescopes (FPMA and FPMB). 
For J0823, we limit the modeling to FPMB, since the source is only
fully within the \nustar FoV for FPMB.

For the six sources with soft X-ray counterparts, we extract
additional spectra from the archival soft X-ray
datasets detailed in Table \ref{xrayData_table}, using the relevant
software packages (the \chandra Interactive Analysis Observations
software,\footnote{\citet{Fruscione06}; http://cxc.harvard.edu/ciao/index.html} the
\swiftxrt analysis software distributed with
HEASoft,\footnote{http://www.swift.ac.uk/analysis/xrt/} and the \xmm
Science Analysis Software\footnote{http://xmm.esa.int/sas/}).
We adopt source extraction apertures of $5''$, $10''$, 
and $12$--$15''$
radius for the \chandra, \swiftxrt, and \xmm data, respectively. 
For J1444 we coadd the \swiftxrt spectra across all six observations,
since the source is only significantly detected in the coadded data.

%%%%%%%%%%%%%%%%%%%%%%%%%%%%%%%%%%%%%%%%%%%%%%%%%%%%%%%%%%%%%%%%%%%%%%
\section{X-ray properties}
\label{xray}
%%%%%%%%%%%%%%%%%%%%%%%%%%%%%%%%%%%%%%%%%%%%%%%%%%%%%%%%%%%%%%%%%%%%%%

\subsection{X-ray spectral modeling}
\label{xray_modelling}

We perform X-ray spectral modeling using \xspec (version 12.8.1j;
\citealt{Arnaud96}) with the $C$~statistic ($\mathtt{cstat}$)
setting,\footnote{The $W$~statistic is actually used, since the
  background is unmodelled; see http://heasarc.gsfc.nasa.gov/docs/xanadu/xspec/wstat.ps.} 
which is more appropriate than $\chi^{2}$ in the low-counts regime
\citep[e.g.,][]{Nousek89}. We group the data (source plus background)
from \nustar and from other X-ray missions by a minimum of $3$ counts
and $1$ count per bin, respectively, as recommended for use with
$\mathtt{cstat}$.\footnote{https://asd.gsfc.nasa.gov/XSPECwiki/low\_count\_spectra}

\begin{figure*}
\centering
\includegraphics[width=1.0\textwidth]{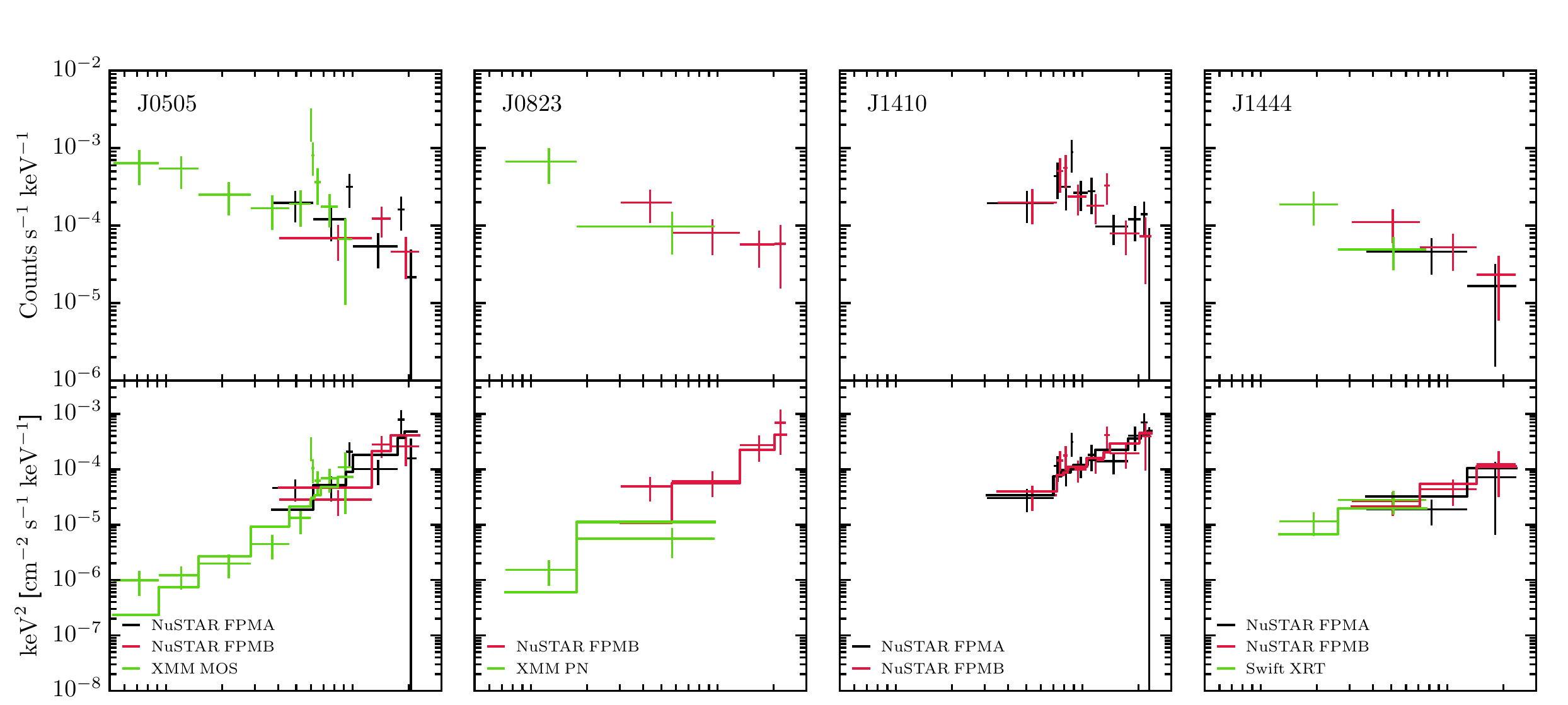}
\includegraphics[width=1.0\textwidth]{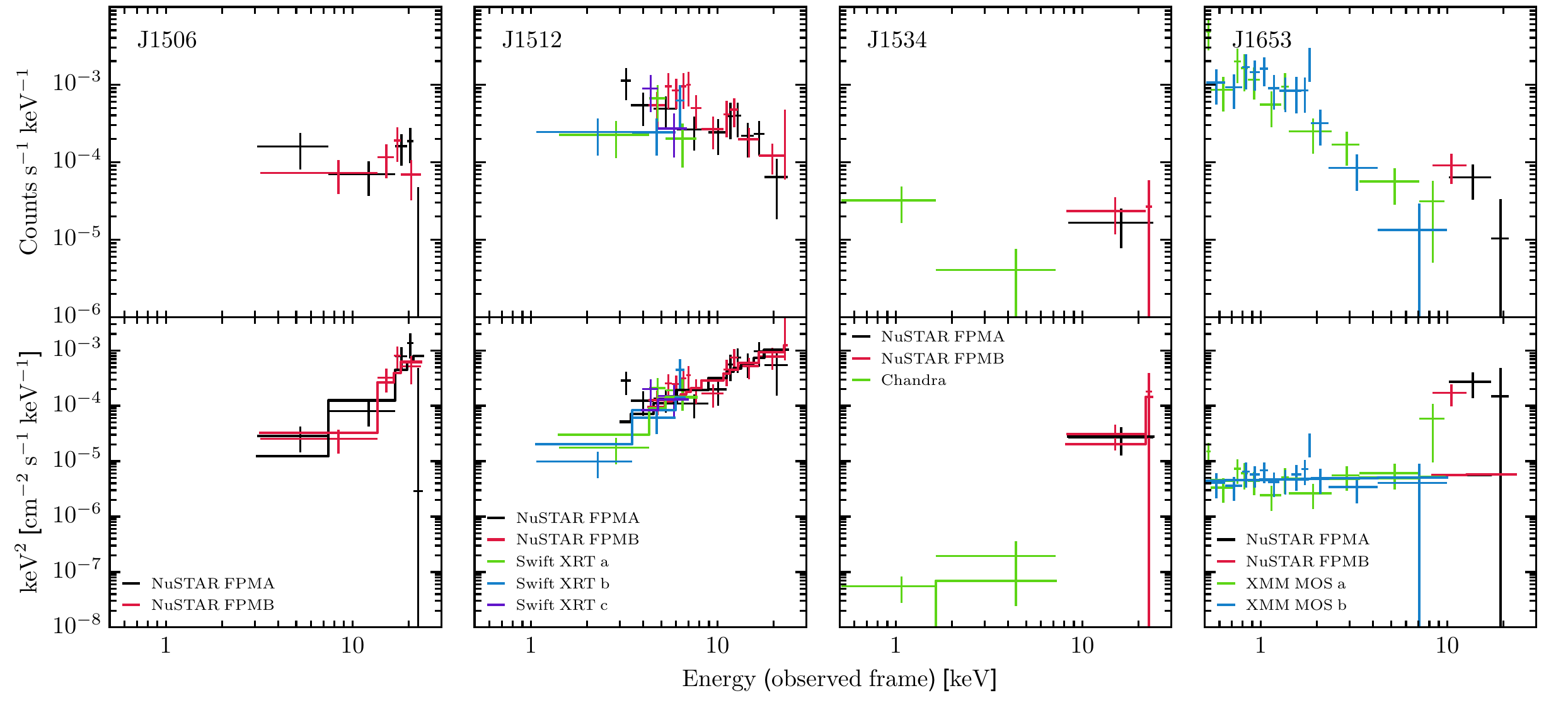}
\caption{X-ray spectra in observed count-rate units (top panel for a
  given source)
  and in $EF_{E}$ units (bottom panel for a given source) for the eight extreme
  \nustar sources (Section \ref{xray}). Black and red
  correspond to \nustar FPMA and FPMB, respectively. The green, blue, and purple spectra 
  represent the available soft X-ray data (as labelled). Letter
  suffixes (e.g., \swiftxrt~b) indicate separate observations. See
  Table \ref{xrayData_table} for a full description of the adopted
  datasets.
  The data are binned to a mimimum
  significance of $2\sigma$ per bin for visual purposes.
  The $EF_{E}$ spectra are shown with best-fitting power law models,
  binned to match the data (solid lines). 
}
\label{eeuf} 
\end{figure*}

In all cases, we fit a simple unabsorbed power law model
in order to constrain the effective photon index ($\Gamma_{\rm eff}$),
and thus obtain a basic measure of the overall X-ray spectral slope.
Figure \ref{eeuf} shows the \nustar plus soft X-ray (\chandra,
\swiftxrt, or \xmm) spectra for the eight extreme \nustar
serendipitous survey sources, with power law model fits to each.
Flat $\Gamma_{\rm eff}$ values (e.g., $\lesssim 0.5$) give empirical evidence for
high or CT absorption. Further empirical evidence for CT absorption can be obtained from
the detection of a strong fluorescent \feka emission line at $\approx 6.4$~keV (with
an equivalent width of $\mathrm{EW}_{\rm Fe\ K\alpha}>1$~keV, although
lower values do not necessarily rule out CT absorption; e.g., \citealt{DellaCeca08,Gandhi16}). This reflection
feature becomes more prominent with increasing levels of
absorption \citep[e.g.,][]{Risaliti02}. 
To place constraints on $\mathrm{EW}_{\rm Fe\ K\alpha}$ for our sources, we model
the rest-frame $\approx 4$--$9$~keV spectrum as a
power law (to fit the continuum) plus an unresolved
Gaussian at rest-frame $E=6.4$~keV. For two sources (J0505 and J1512)
the emission line is well detected, and $\mathrm{EW}_{\rm Fe\
  K\alpha}$ is therefore constrained. 
For the remaining six sources the line is undetected, due to
insufficient counts, and we report upper limits on $\mathrm{EW}_{\rm
  Fe\ K\alpha}$ (assuming a line width of $\sigma_{\rm line}=0.1$~keV) where the data allow
informative constraints.
In Table \ref{xrayBasic_table} we
provide the basic observed X-ray spectral properties for the sample:
effective photon indices, \feka line equivalent widths, and observed
(i.e., uncorrected for absorption) X-ray luminosities. 
\renewcommand*{\arraystretch}{1.1}
\begin{table}
\centering
\caption{Basic X-ray spectral parameters}
\begin{tabular}{lccccc} 
\hline\hline \noalign{\smallskip}
Object & $\Gamma^{\mathrm{NuSTAR}}_{\rm eff}$ &
                                            $\Gamma^{\mathrm{soft}}_{\rm eff}$ & $\mathrm{EW}_{\rm FeK\alpha}$ & $L^{\rm obs}_{\rm 2-10}$ & $L^{\rm obs}_{\rm 10-40}$ \\
(1) & (2) & (3) & (4)  & (5)  & (6) \\
\noalign{\smallskip} \hline \noalign{\smallskip}
J0505 & $-0.1^{+0.7}_{-0.8}$ & $-0.9^{+0.8}_{-1.4}$ & $1.4^{+1.4}_{-0.9}$ & $41.3$ & $42.3$ \\
J0823 & $0.3^{+1.1}_{-1.3}$ & $1.2^{+1.2}_{-0.9}$ & $\cdots$ & $42.5$ & $44.4$ \\
J1410 & $0.3\pm0.4$ & $\cdots$ & $<1.7$ & $42.0$ & $42.7$ \\
J1444 & $-0.3^{+0.9}_{-1.2}$ & $0.7\pm1.1$ & $<1.4$ & $44.7$ & $45.1$ \\
J1506 & $-0.7^{+0.9}_{-1.6}$ & $\cdots$ & $<3.2$ & $39.9$ & $42.6$ \\
J1512 & $0.9^{+0.4}_{-0.5}$ & $-0.6^{+0.7}_{-0.9}$ & $0.76^{+1.04}_{-0.56}$ & $42.4$ & $43.2$ \\
J1534 & $<-0.9^{~\dagger}$ & $3.3^{+5.9}_{-2.4}$ & $\cdots$ & $39.8$ & $42.7$ \\
J1653 & $-0.5^{+0.9~\dagger}_{-0.6}$ & $2.0\pm0.3$ & $<0.5$ & $42.7$ & $44.3$ \\
\noalign{\smallskip} \hline \noalign{\smallskip}
\end{tabular}
\begin{minipage}[c]{0.48\textwidth}
\footnotesize
\textbf{Notes.} (1): Abbreviated \nustar source name. (2):  The
\nustar effective photon index; i.e., the photon index
obtained from approximating the \nustar $3$--$24$~keV spectrum as a
simple power law. For the sources marked $^{\dagger}$, the
constraint was obtained using a combination of \nustar and soft X-ray
(\xmm or \swiftxrt) data, due to weak \nustar-only constraints. (3): The
``soft'' effective photon index, measured using the
available soft X-ray spectra from \chandra, \swiftxrt, or \xmm (over
the full energy range for the relevant observatory; $\approx
0.5$--$10$~keV). 
(4): Constraint on the \feka line
equivalent width (\ewfeka). Units: $\mathrm{keV}$. (5) and (6): Logarithm of the observed (i.e., uncorrected for
absorption) X-ray luminosities in the rest-frame $2$--$10$~keV and
$10$--$40$~keV bands, respectively. Units: \ergpersec.
\end{minipage}
\label{xrayBasic_table}
\end{table}

We use three more 
spectral models in order to constrain the
source properties such as the intrinsic absorbing column density (\nh),
the intrinsic photon index ($\Gamma$), and the X-ray luminosity.
Firstly, we fit a transmission-only model (the \transmission model,
hereafter): a power law attenuated by
redshifted photoelectric absorption and Compton scattering of photons
out of the
line of sight ($\mathtt{CABS\cdot ZWABS\cdot POW}$, in \xspec
formalism). This model represents one extreme of obscured AGN spectra,
where the X-ray spectrum is dominated by the primary AGN continuum transmitted directly
along the line of sight. Secondly, we fit a reflection-only model (the \reflection
model, hereafter), which represents a power law spectrum reflected by
circumnuclear material. For this we use the \pexrav model \citep{Magdziarz95} with
the reflection scaling factor set to $-1$ to yield a pure reflection
spectrum, and with the other parameters set to default values. This
model represents the other extreme of obscured AGN spectra, where the
X-ray spectrum is dominated by the reflected AGN continuum, which (in
combination with strong Fe line emission) implies very high
column densities ($N_{\rm H}\gg 10^{24}$~\nhunit). 
At high column densities,
X-ray spectra are typically more complex than the \transmission and
\reflection models above, and ideally any absorbed continuum,
reflected continuum, and fluorescent line emission should be modeled
in a self-consistent way and assuming a well-motivated
geometry. We therefore perform an additional third test using the
\bntorus model (the \torus model, hereafter; \citealt{Brightman11}),
which was produced using simulations of X-ray radiative
transfer through a toroidal distribution of gas. We set the model to
an edge-on torus configuration (with $\theta_{\rm
  inclination}$ and $\theta_{\rm torus}$ set to $87$\degrees and $60$\degrees, respectively). In this form, the
\torus model has the same number of free parameters as the
\transmission and \reflection models, and is therefore no less suited to the
statistical quality of the data.
For every model fit, we account for Galactic
absorption with a $\mathtt{PHABS}$ multiplicative component, fixed to
column density values from \citet{Kalberla05}. 
In cases where $\Gamma$ and \nh cannot be simultaneously constrained,
we fix the intrinsic photon index at $\Gamma=1.9$ (a typical value for
AGNs detected at $3$--$24$~keV; e.g., \citealt{Alexander13}; Del~Moro
et al.\ 2017, submitted).
In Table
\ref{xrayModels_table} we show the best-fit parameters obtained by
applying the three models described above:
intrinsic photon indices, column densities, fit statistics,
and intrinsic (i.e., absorption-corrected) luminosities. 
\renewcommand*{\arraystretch}{1.1}
\begin{table*}
\centering
\caption{Best-fit parameters for the X-ray spectral modeling}
\footnotesize\addtolength{\tabcolsep}{-2.7pt}
\begin{tabular}{lc|cc|ccc|cc|ccc|ccc} 
\hline\hline \noalign{\smallskip}
\multicolumn{2}{l}{} & \multicolumn{2}{c}{$\mathtt{pow}$} & \multicolumn{3}{c}{\transmission} &
\multicolumn{2}{c}{\reflection} & \multicolumn{3}{c}{\torus} & \multicolumn{3}{l}{} \\
\cmidrule(rl){3-4} \cmidrule(rl){5-7} \cmidrule(rl){8-9} \cmidrule(rl){10-12} 
Object & $E$ range & $\Gamma_{\rm eff}$ & $C/n$ & $\Gamma$ & \nh &
                                                                   $C/n$ & $\Gamma$ & $C/n$
  & $\Gamma$ & \nh & $C/n$ & $L^{\rm int}_{\rm 2-10}$ & 
$L^{\rm int}_{\rm 10-40}$ & CT \\
  & (keV) &  &  &  & ($10^{24}$~\nhunit) &  &  &  &  & ($10^{24}$~\nhunit) &  & &  & \\
(1) & (2) & (3) & (4)  & (5)  & (6) & (7) & (8) & (9) & (10)  & (11)
  & (12) & (13) & (14) & (15) \\
\noalign{\smallskip} \hline \noalign{\smallskip}
J0505 & $0.5$--$24$ & $-0.2\pm0.2$ & $164/142$ & $[1.9]$ & $0.87^{+0.37}_{-0.27}$ & $159/139$ & $1.3\pm0.4$ & $148/139$ & $2.5^{+0.4}_{-0.8}$ & $1.5^{+4.7}_{-0.5}$ & $148/142$ & $43.1$ & $42.7$ & Y \\
J0823 & $0.5$--$24$ & $-0.2\pm0.7$ & $78/54$ & $[1.9]$ & $0.73^{+1.51}_{-0.61}$ & $45/33^{\dagger}$ & $2.6^{+1.0}_{-0.7}$ & $71/53$ & $[1.9]$ & $12.6^{+\mathrm{u}}_{-12.0}$ & $41/33^{\dagger}$ & $44.4$ & $44.4$ & $\cdots$ \\
J1410 & $3$--$24$ & $0.3\pm0.4$ & $78/87$ & $[1.9]$ & $0.74^{+0.31}_{-0.25}$ & $78/87$ & $1.8\pm0.4$ & $82/87$ & $[1.9]$ & $0.63^{+0.31~*}_{-0.24}$ & $80/87$ & $\cdots$ & $43.0$ & $\cdots$ \\
J1444 & $0.6$--$24$ & $0.8\pm0.5$ & $98/75$ & $[1.9]$ & $0.21^{+0.28}_{-0.17}$ & $104/75$ & $2.1^{+0.7}_{-0.6}$ & $102/75$ & $[1.9]$ & $0.21^{+0.28~*}_{-0.17}$ & $103/75$ & $45.1$ & $45.1$ & $\cdots$ \\
J1506 & $3$--$24$ & $-0.7^{+0.9}_{-1.6}$ & $77/64$ & $[1.9]$ & $5.0^{+3.6}_{-3.7}$ & $82/64$ & $[1.9]$ & $79/65$ & $1.5^{+1.2}_{-\mathrm{u}}$ & $4.1^{+\mathrm{u}}_{-2.3}$ & $70/63$ & $\cdots$ & $43.3$ & Y \\
J1512 & $0.6$--$24$ & $0.4\pm0.2$ & $123/98$ & $[1.9]$ & $0.13^{+0.22}_{-0.06}$ & $142/98$ & $2.1^{+0.2}_{-0.3}$ & $112/98$ & $2.8^{+\mathrm{u}~**}_{-0.8}$ & $2.9^{+\mathrm{u}}_{-1.2}$ & $112/97$ & $44.6$ & $44.0$ & Y \\
J1534 & $0.5$--$24$ & $-2.3^{+1.5}_{-\mathrm{u}}$ & $90/74$ & $[1.9]$ & $2.5^{+\mathrm{u}}_{-1.2}$ & $84/72$ & $[1.9]$ & $90/73$ & $[1.9]$ & $1.6^{+\mathrm{u}}_{-1.1}$ & $87/72$ & $42.7$ & $42.7$ & y \\
J1653 & $0.5$--$24$ & $1.9^{+0.4}_{-0.3}$ & $182/194$ & $2.3^{+0.5}_{-0.4}$ & $2.4^{+1.3}_{-0.9}$ & $165/192$ & $2.4^{+0.8}_{-0.5}$ & $179/193$ & $2.3^{+0.6}_{-0.5}$ & $1.6^{+1.5}_{-1.1}$ & $175/192$ & $44.3$ & $44.1$ & y? \\
\noalign{\smallskip} \hline \noalign{\smallskip}
\end{tabular}
\begin{minipage}[c]{0.97\textwidth}
\footnotesize
\textbf{Notes.} (1): Abbreviated \nustar source name. (2): Energy
range modeled (units of keV). 
(3)--(4): Best-fit results for the unobscured power law model ($\mathtt{pow}$; also shown
in Figure \ref{eeuf}), where
$\Gamma_{\rm eff}$ is the power law photon index.
(5)--(12): Best-fit results for the
\transmission, \reflection, and \torus models, respectively. These
include the intrinsic photon index ($\Gamma$; square brackets indicate
fixed values), the column density
(\nh; units of $10^{24}$~\nhunit), and the fit statistic ($C/n$, where $C$ is the C-statistic
and $n$ is the number of degrees of freedom). An error value of
$\mathrm{+u}$ or $\mathrm{-u}$ indicates that the parameter is
unconstrained at the upper or lower end. (13) and (14): Logarithm of the intrinsic (i.e.,
absorption-corrected) X-ray luminosities in the rest-frame $2$--$10$~keV and
$10$--$40$~keV bands, respectively. Units: \ergpersec. 
(15): Flag to indicate high-confidence CT AGNs and likely-CT AGNs (marked as
``Y'' and ``y'', respectively). J1653 is marked as ``y?'' since there
is multiwavelength evidence against a CT interpretation (Section
\ref{indirect}). For the three sources marked as ``$\cdots$'', we
cannot strongly rule out CT absorption based on the X-ray modeling.
$^{*}$: For two
sources (J1410 and J1444) we show the conservative low-\nh
\torus model solution in this Table, but in each case there is also a second
similarly valid solution
at very high column densities (for J1410, $N_{\rm H} >6\times
10^{24}$~\nhunit and $C/n =92/87$; and for J1444, $N_{\rm
  H}>6\times 10^{24}$~\nhunit and $C/n=102/75$). 
$^{**}$: For J1512, fixing $\Gamma$ to more typical
values results in even higher-\nh solutions (e.g., a lower limit of
$N_{\rm H}>8\times 10^{24}$~\nhunit for $\Gamma=1.9$).
$^{\dagger}$: As detailed in Section \ref{xray_modelling}, the 
\transmission and \torus model fits for J0823 are performed for
the \nustar data only (i.e., the \xmm data are excluded).
\end{minipage}
\label{xrayModels_table}
\end{table*}

In one case (J1653) we find that an additional soft X-ray dominated
model component is necessary 
to obtain an acceptable fit to the data. 
For J1653 all three models provide a poor fit to the \xmm plus \nustar
spectrum (for the \transmission, \reflection, and \torus models, the
ratio of the $C$~statistic to the number of degrees of freedom is $C/n
= 352/200$, $311/202$, and $335/201$, respectively) and leave
strong positive residuals at high energies ($\gtrsim 8$~keV). 
This is due to an apparently sudden change in the spectral shape, with
the low energies ($\lesssim 4$~keV) dominated by a steep ($\Gamma
\approx 2$) component and the higher energies ($\gtrsim 4$~keV)
dominated by a flatter component ($\Gamma\approx -0.5$). 
One way to interpret this is an electron-scattered or leaked (due to
partial covering) AGN power law at
lower energies and a primary AGN continuum penetrating through at higher
energies, as is commonly observed for well-studied AGNs in the local
Universe (e.g., \citealt{Cappi06}). The relatively high luminosity ($L_{\mathrm{0.5-4\ keV}}\approx 7\times
10^{42}$~\ergpersec) 
justifies the scattered AGN
power law interpretation rather than, e.g., thermal emission
associated with star formation. For J1653 we therefore add an unobscured power law
component to the three spectral models, with the spectral slope tied
to that of the intrinsic AGN power law continuum. This results in
statistically improved fits (see the $C/n$ values in Table \ref{xrayModels_table}), and reasonable scattered power law
fraction constraints ($f_{\rm scatt}\approx 0.04$--$5\%$). 

The source J1534 also shows evidence for a steep soft
  component in the \chandra spectrum ($\Gamma_{\rm eff}\approx 3$ at
  $0.5$--$8$~keV), which is dominated by photon counts at $<2$~keV (as
  described in Section \ref{xray_cparts}).
This is uncharacteristic of pure AGN emission and indicates
that at low X-ray energies there is a significant contribution to the
spectrum from other radiative
processes in the host galaxy. We find that the detection of this soft
component is due to the primary AGN spectrum being highly absorbed (see Sections
\ref{xray_results} and \ref{indirect}) so as not to be
well detected by \chandra. Indeed, the AGN is only detectable at
$>8$~keV with \nustar.
The luminosity of the soft X-ray emission ($L^{\rm obs}_{\rm
  2\mbox{-}10}=10^{39.8}$~\ergpersec; Table \ref{xrayBasic_table}) is
in broad agreement with the
expectations for normal galaxy emission based on the X-ray main
sequence of star formation (\citealt{Aird17a}) and given the
stellar mass of J1534 ($M_{\star}=10^{11.1}\,M_{\odot}$; based on the SED modeling in
Section \ref{indirect}). If the soft component is instead interpreted
as a scattered AGN power law, then the scattered fraction must 
be small ($f_{\rm scatt}\lesssim 0.05\%$). 
For the spectral modeling of J1534 below, we parameterize the steep soft
emission with an additional power law component. We also tested a
different approach of simply excluding the $<2$~keV photons, and this
yields consistent values for the intrinsic source properties.

For the sources where we model the \nustar data simultaneously with
soft X-ray (\chandra, \swiftxrt, or \xmm) data, there is a general caveat that
the soft X-ray observations are not contemporaneous with the \nustar
data, and AGN variability could thus affect the interpretations. 
Although highly obscured AGNs such as those presented here show
some evidence for lower variability compared to unobscured AGNs (e.g.,
\citealt{Awaki06}), significant variability on
year-long timescales is still possible (e.g., \citealt{Yang16,Masini17}).
While our sources generally show no evidence for significant
variability (e.g., see the overlapping data in Figure \ref{eeuf}), the spectral uncertainties
are generally too large to rule out low-level (e.g., factors of
$\lesssim 2$) variability.
We thus fix the cross-normalization constants to standard values: $1.0$
for \chandra:\nustar; $1.0$ for \swiftxrt:\nustar; and $0.93$ for
\xmm:\nustar (e.g., \citealt{Madsen15}).
There is one exception, J0823, where the \xmm:\nustar
cross-normalization parameter must be left free to obtain
statistically acceptable solutions. The \transmission and
\torus models converge to extremely low cross-normalization constants
($\approx 0.01$), and we therefore limit the modeling to the \nustar
data only. The best-fit \reflection model, however, has a
less extreme cross-normalization constant of
$0.12^{+0.19}_{-0.08}$ when fitting the \xmm plus \nustar data
set. The low cross-normalization constants for J0823 may be
due to X-ray variability between the 2007 \xmm and the
2014 \nustar observations, although we do not draw strong conclusions
given the uncertainties for this source.

\subsection{Results for the X-ray source properties}
\label{xray_results}

\begin{figure}
\centering
\includegraphics[width=0.47\textwidth]{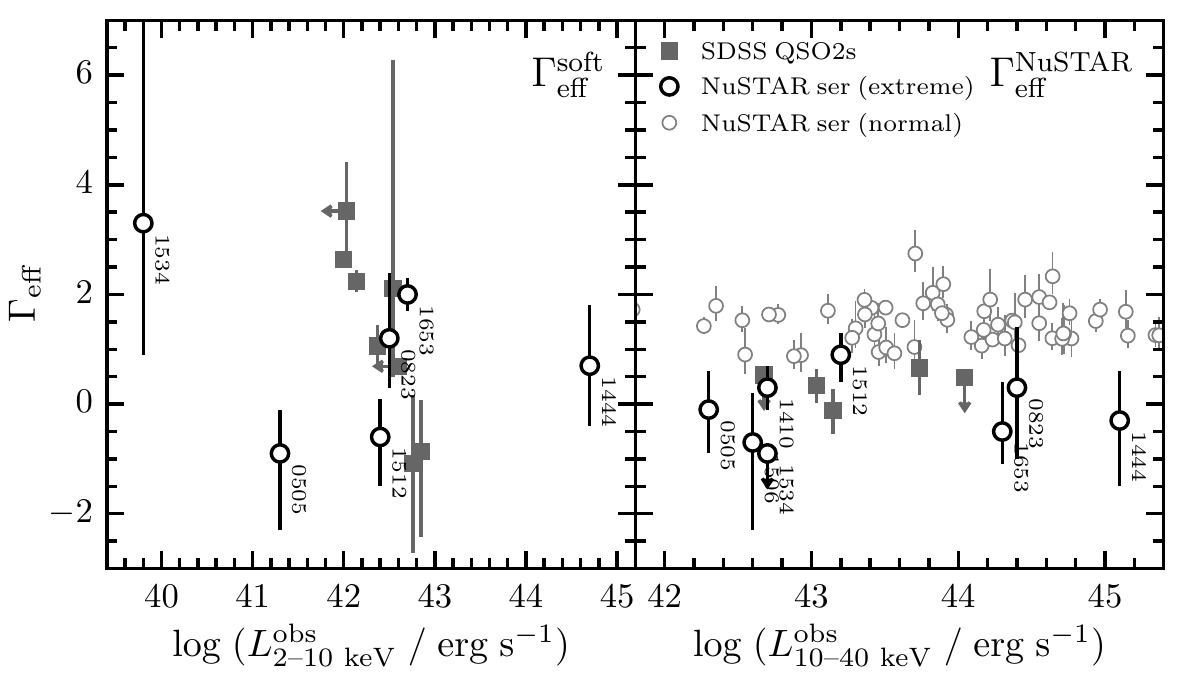}
\caption{Observed X-ray properties: effective photon index (i.e.,
  spectral slope) versus rest-frame X-ray luminosity (uncorrected for
  absorption). The left panel shows the properties measured at
  soft X-ray energies (with \chandra, \swiftxrt, or \xmm), and the
  right panel shows the properties measured at harder X-ray energies
  with \nustar. $\Gamma_{\rm eff}^{\rm soft}$ and $\Gamma_{\rm
    eff}^{\rm NuSTAR}$ are measured for the observed-frame $\approx 0.5$--$10$~keV
  and $3$--$24$~keV bands, respectively.
  We compare the extreme \nustar serendipitous survey
  sources (black circles, individually labelled) to ``normal''
  serendipitous survey sources (smaller grey circles) and to highly
  obscured and CT Type 2 quasars which were optically selected and followed
  up with \nustar observations (filled gray squares; \citealt{Gandhi14,Lansbury14,Lansbury15}).
}
\label{gam_lx} 
\end{figure}

Here we summarize the measured X-ray properties.
Figure~\ref{gam_lx} shows the effective photon indices (i.e., the
observed spectral slopes) of the sources, as measured with individual
X-ray observatories, as a function of 
X-ray luminosity (uncorrected for absorption). 
The extreme \nustar sources cover a broad range in luminosity. 
The \nustar-measured effective photon indices (right panel of Figure
\ref{gam_lx}) are generally very low (median value of $\Gamma_{\rm
  eff}=-0.2$ at $3$--$24$~keV), giving
empirical evidence for very high absorption
levels. We compare against another sample of extreme systems: highly
obscured SDSS-selected Type 2 quasars targeted with \nustar
(\citealt{Gandhi14,Lansbury14,Lansbury15}). The two extreme samples
cover a similar range of spectral slopes, and lie at significantly harder
values (i.e. lower $\Gamma_{\rm eff}$ values) than the
general population of ``normal'' \nustar
serendipitous survey sources (also shown in Figure \ref{gam_lx}, for
sources with constrained $\Gamma_{\rm eff}$ values;
\citealt{Lansbury17}). 
The measured spectral slopes show a large scatter at soft energies ($\approx
0.5$--$10$~keV; using \chandra, \swiftxrt, and \xmm). For the
\nustar-observed SDSS Type 2 quasars, this scatter was found to be partly due to an
increased contamination at these lower X-ray energies from radiative processes
other than the direct AGN emission (e.g., \citealt{Lansbury15}), which
may also be the case for some of the extreme \nustar sources (namely
J1534 and J1653; see Section \ref{xray_modelling}). In other words,
soft X-ray observations alone would fail to identify
  $57^{+19}_{-21}\%$ of the extreme sources in Figure
\ref{gam_lx} as highly obscured using spectral slope information
(assuming a threshold of $\Gamma_{\rm eff}=1$).
\nustar observations on the other hand are highly reliable at
identifying the most highly obscured AGNs.

\begin{figure}
\centering
\includegraphics[width=0.47\textwidth]{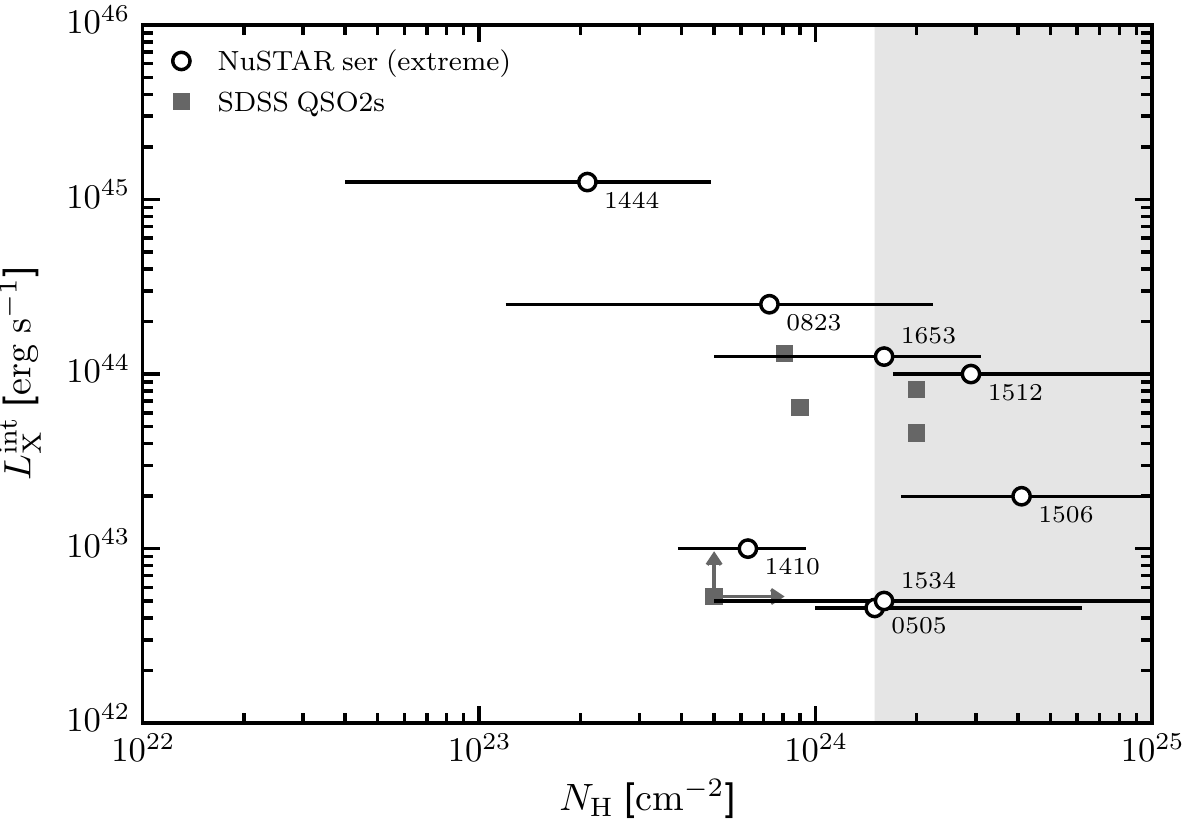}
\caption{Rest-frame intrinsic (i.e., absorption-corrected) $10$--$40$~keV X-ray luminosity
  (\lx) versus column density (\nh), from modeling the X-ray spectra
  of the extreme \nustar serendipitous survey sources (open circles). Each data point 
  corresponds to the \torus model solution (except J0823, where the \transmission model
  solution is shown).
  Following Figure \ref{gam_lx}, the filled gray
  squares show a comparison sample of highly obscured Type 2 quasars
  (\citealt{Gandhi14,Lansbury14,Lansbury15}). The CT column density
  region ($N_{\rm H}\ge 1.5\times 10^{24}$~\nhunit) is highlighted in gray.
}
\label{lx_nh} 
\end{figure}

For the purposes of comparing \nh constraints and estimating intrinsic
luminosities (\lx; shown in Table \ref{xrayModels_table}), we adopt
the \torus model solutions. In one exception (J0823) we adopt the
lower-\nh \transmission model solution.
The adopted best-fitting \nh and \lx values are shown in
Figure \ref{lx_nh}. 
Based on these intrinsic luminosity constraints, the more distant AGNs
($z>0.2$) are at ``X-ray quasar'' luminosities
($L_{\rm X}\gtrsim 10^{44}$~\ergpersec), and the less distant AGNs
($z<0.2$) range from relatively low luminosities up to the quasar
threshold ($L_{\rm X}\approx 10^{42.7}$--$10^{44}$~\ergpersec).
The \nh constraints shown may be
conservative for sources where the \reflection model gives a
statistically acceptable fit to the X-ray spectrum (indicating consistency with $N_{\rm
  H}\gg 10^{24}$~\nhunit).
For a similar reason, the Compton-thin constraints shown for J1410 and
J1444 may be conservative; the \torus modeling also finds 
statistically acceptable reflection-dominated model solutions at very high, CT column densities
($N_{\rm H}>6\times 10^{24}$~\nhunit) in these cases. Nevertheless,
for these two sources we assume the lower-\nh, Compton-thin solutions on the
basis that their X-ray to MIR luminosity ratios are consistent with those
for unobscured AGNs (Section \ref{indirect}).

Considering all of the X-ray spectral constraints together, there are
three sources with strong evidence for being CT AGNs (J0505, J1506,
and J1512; two of which have supporting evidence from high
equivalent width \feka emission, as shown in Table
\ref{xrayBasic_table}), 
one likely-CT AGN (J1534; supporting indirect evidence is
presented in Section \ref{indirect}), 
one possible CT AGN (J1653; although the indirect evidence prefers a
lower-obscuration solution; see Section \ref{indirect}), 
one highly obscured Compton-thin AGN (J1410), 
one uncertain but likely highly obscured AGN (J0823),
and one likely moderately absorbed AGN (J1444).
Of the total four likely-CT AGNs identified with \nustar, none would be identified as CT using
just the soft X-ray ($<10$~keV) data, except possibly J0505 for which the \xmm
spectrum alone shows good evidence for a $\gtrsim 1$~keV \feka line.

Prior to this work, only one other AGN has been identified in the
\nustar extragalactic surveys with strong evidence for CT absorption.
This source, ID~330, was identified in the \nustar-COSMOS survey
(\citealt{Civano15}; Zappacosta et al.\ 2017, submitted). Like the robust CT AGNs
presented here (J0505, J1506, and J1512), ID~330 lies at low redshift
($z=0.044$), and has a high \nustar band ratio (see Figure \ref{br_z}). 
Assuming a \bntorus-based model to fit the
X-ray spectrum, the column density of ID~330 is $N_{\rm
  H}=(1.2^{+0.3}_{-0.1})\times 10^{24}$~\nhunit (\citealt{Civano15}),
which is similar to J0505 and less extreme than J1506 and J1512.
Additional CT candidates are identified by Del Moro et al.\
(2017, submitted) and Zappacosta et al.\ (2017, submitted), as part of
studies which focus on the broad X-ray spectral properties of \nustar 
extragalactic survey sources. 
We note that our extreme sample (selected from the total
$40$-month serendipitous catalog; see Section \ref{selection}) does not overlap with
the Zappacosta et al.\ (2017, in prep.)\ sample, which is a subset of $24$ serendipitous
sources (plus $39$ sources from the \nustar dedicated-field surveys).

%%%%%%%%%%%%%%%%%%%%%%%%%%%%%%%%%%%%%%%%%%%%%%%%%%%%%%%%%%%%%%%%%%%%%%
\section{Indirect Absorption Diagnostics}
\label{indirect}
%%%%%%%%%%%%%%%%%%%%%%%%%%%%%%%%%%%%%%%%%%%%%%%%%%%%%%%%%%%%%%%%%%%%%%

The intrinsic X-ray and MIR luminosities of AGNs are
tightly correlated
\citep[e.g.,][]{Krabbe01,Lutz04,Horst08,Fiore09,Gandhi09,Lanzuisi09,Ichikawa12,Matsuta12,Asmus15,Mateos15,Stern15,Chen17}.
The observed X-ray to MIR luminosity ratio of a source can therefore give an
independent, albeit indirect, assessment of the degree of obscuration
(e.g., see \citealt{Alexander16} for a recent review); the \textit{observed} X-ray
luminosity for any significantly absorbed AGN will be suppressed with
respect to the \textit{intrinsic} luminosity, causing it to deviate
from the X-ray to MIR luminosity relation. This diagnostic has been utilized for
other \nustar studies of obscured AGNs
\citep[e.g.,][]{Balokovic14,Lansbury14,Stern14,Annuar15,Lansbury15,Gandhi16,LaMassa16,Annuar17}.

\begin{figure*}
\centering
\includegraphics[width=1.0\textwidth]{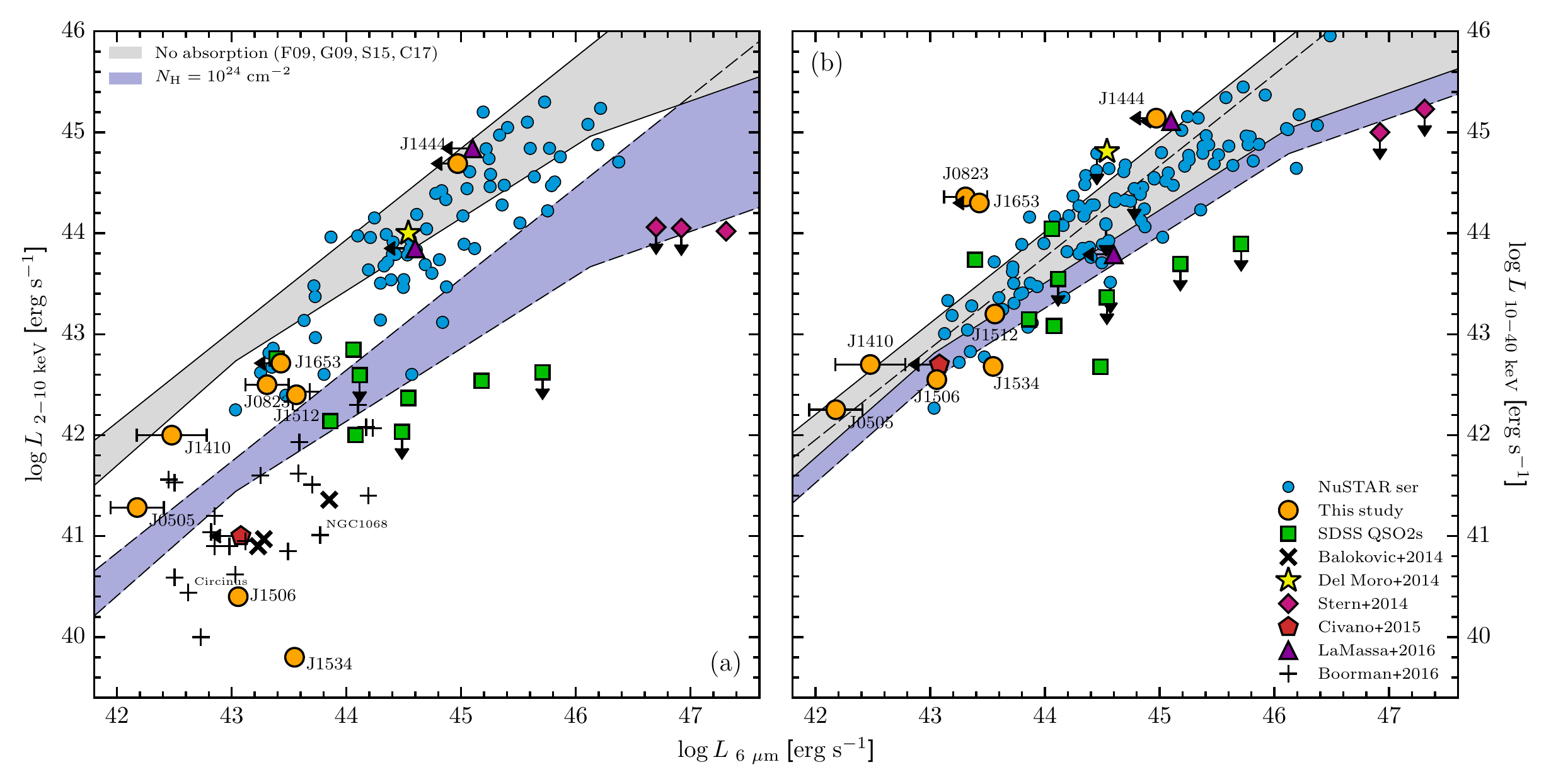}
\caption{X-ray luminosities (at rest-frame $2$--$10$~keV and
  $10$--$40$~keV) versus rest-frame \sixum luminosity in $\nu L_{\rm
    \nu}$ units (\Lsixum). For the data points,
  we show observed X-ray luminosities (i.e., uncorrected for line-of-sight absorption
  of the X-rays). 
  The extreme \nustar serendipitous survey sources are highlighted as
  orange circles, and are individually labeled.
 We compare to ``normal'' \nustar serendipitous survey sources (smaller
 blue circles; \citealt{Lansbury17}) and to other \nustar-observed samples of obscured to CT AGNs (see
 figure legend). 
 We also compare with known ``bona fide'' CT
  AGNs in the local universe (``$+$'' symbols; $\mathrm{distance}\lesssim 100$~Mpc;
  data compiled in Boorman et al.\ 2017, in prep.), including NGC~1068
  and Circinus. 
 The gray regions (with solid borders) highlight the range of luminosity ratios expected in
 the case of zero X-ray absorption (based on \citealt{Gandhi09,Fiore09,Stern15,Chen17}), and
 the purple regions (with dashed borders) show the approximate X-ray suppression expected for absorption
 by gas with a column density of $N_{\rm H}=10^{24}$~\nhunit.}
\label{lx_lmir}
\end{figure*}

Figure \ref{lx_lmir} shows the observed X-ray versus intrinsic \sixum
luminosities for the eight extreme \nustar serendipitous survey sources.
Adopting the methodology of \citet{Assef08,Assef10,Assef13}, the AGN
\Lsixum values have been determined using SED modeling of the SDSS and \wise
photometry available, where each SED is modeled as the
best-fit linear combination of four empirical templates (one AGN template and
three different galaxy templates; \citealt{Assef10}). 
The approach allows constraints on the relative contribution of the AGN and the
host galaxy to the observed luminosity (see 
\citealt{Lansbury14,Lansbury15} for applications of the same technique
to an SDSS \typeii quasar sample).
For two of the extreme \nustar sources (J1444 and J1653) the SED
modeling results are consistent with zero
contribution from the AGN, and we therefore adopt
conservative upper limits for \Lsixum (Figure \ref{lx_lmir}). 
For the remaining six sources, the AGN contributes between $\approx
0.07$ and $\approx 0.77$ of
the overall luminosity, for the $0.1$--$30$~$\mathrm{\mu m}$ wavelength
range (see Table \ref{SED_table}). The resulting uncertainties on
\Lsixum (also listed in Table \ref{SED_table}) are determined 
from a Monte Carlo re-sampling of the photometric data over $1000$
iterations, and are shown in Figure \ref{lx_lmir}.

\begin{table}
\centering
\caption{SED modeling results}
\begin{tabular}{ccc} \hline\hline
  \noalign{\smallskip}
Object  & $\hat{a}$ & \Lsixum \\
     &      &   $10^{42}$~\ergpersec \\
(1)  & (2) & (3) \\
\noalign{\smallskip} \hline \noalign{\smallskip}
J0505  & $0.07\pm0.05$ & $1.5\pm0.8$ \\
J0823  & $0.28\pm0.08$ & $20.3\pm8.8$ \\
J1410  & $0.11\pm0.07$ & $3.0\pm2.1$ \\
J1444  & $0.00^{+0.19}$ & $<933.2$\\
J1506  & $0.28\pm0.01$ & $11.4\pm0.7$ \\
J1512  & $0.76\pm0.09$ & $36.6\pm1.7$ \\
J1534  & $0.40\pm0.03$ & $35.3\pm3.8$ \\
J1653  & $0.02^{+0.06}_{-0.02}$ & $<26.8$ \\
\noalign{\smallskip} \hline \noalign{\smallskip}
\end{tabular}
\begin{minipage}[c]{0.26\textwidth}
\footnotesize
\textbf{Notes.} (1): Abbreviated \nustar source name. 
(2): The fractional contribution of the AGN to the intrinsic
luminosity at $0.1$~$\micron$--$30$~$\micron$. 
(3): The rest-frame \sixum luminosity of the AGN.
\end{minipage}
\label{SED_table}
\end{table}

In Figure \ref{lx_lmir} we compare with ``normal'' \nustar
serendipitous survey sources (\citealt{Lansbury17}) and
with other \nustar-observed highly obscured AGNs, including:
nearby CT AGNs identified in the \nustar snapshot survey
($z\approx0.01$; \citealt{Balokovic14}); candidate CT \typeii quasars selected
by SDSS ($z=0.05$--$0.49$; \citealt{Lansbury14,Gandhi14,Lansbury15});
a highly obscured quasar identified in the \nustar-ECDFS survey
($z\approx2$; \citealt{DelMoro14}); and the CT AGN identified in the
\nustar-COSMOS survey ($z=0.044$; C15). Also plotted are ``bona fide''
CT AGNs in the local universe ($\mathrm{distance}\lesssim 100$~Mpc;
data compiled in Boorman et al.\ 2017, in prep.).
We compare all sources
  with the intrinsic X-ray--MIR relation for unobscured AGNs
  \citep{Fiore09,Gandhi09,Stern15,Chen17}, and to demonstrate the expected
  deviation from the relation for highly obscured AGNs, we also show
  the modified relation for X-ray luminosities suppressed by $N_{\rm
    H}=10^{24}$~\nhunit gas. The latter results in a more extreme
  suppression of the X-ray luminosity for the $2$--$10$~keV band (\lx
  is decreased by a factor of $\approx 20$) than for the $10$--$40$~keV
  band (a factor of $\approx 2$ decrease), where the higher energy
  photons are less affected by absorption.

For the eight extreme \nustar serendipitous survey sources, the X-ray
to MIR luminosity ratios are in broad agreement with the X-ray
spectral modeling results, in that the sources with X-ray spectroscopic
evidence for being CT are further offset from the intrinsic \lx--$L_{\rm
MIR}$ relations than the less obscured AGNs. This is especially apparent for J0505, J1506, J1512,
and J1534 at $2$--$10$~keV, where these likely-CT sources overlap well
with the X-ray to MIR luminosity ratios of local ``bona fide'' CT
AGNs, as well as luminous highly obscured and CT Type 2 quasars. 
The \lx--$L_{\rm MIR}$ ratios are very low in the cases of J1506 and
J1534, which appear to lie even lower than local bona fide CT AGNs
(including Circinus and NGC~1068), and have observed X-ray
luminosities which are suppressed by $\approx 2$--$3$ orders of
magnitude.
The X-ray properties of these \nustar sources (Section \ref{xray_results}) suggest that
the X-ray weakness is due to extreme absorption, rather than intrinsic
X-ray weakness (e.g., \citealt{Gallagher01,Wu11,Luo14,Teng15}).
J1653 has a relatively high ratio (at both $2$--$10$~keV and
$10$--$40$~keV), suggesting a low
column density which is in tension with the high value measured
in Section \ref{xray}. 
We note however that not all known CT AGNs have low \lx--$L_{\rm MIR}$
ratios, and a small fraction are even underluminous in MIR emission
compared to the intrinsic
relations (NGC~4945, for instance; e.g., \citealt{Asmus15}), which may in
part result from MIR extinction.
Overall, our indirect analysis does not highlight any additional
likely-CT AGNs in the extreme serendipitous sample which were not
already identified by the X-ray spectral analysis.

%%%%%%%%%%%%%%%%%%%%%%%%%%%%%%%%%%%%%%%%%%%%%%%%%%%%%%%%%%%%%%%%%%%%%%
\section{Optical Properties}
\label{optical}
%%%%%%%%%%%%%%%%%%%%%%%%%%%%%%%%%%%%%%%%%%%%%%%%%%%%%%%%%%%%%%%%%%%%%%

\subsection{Optical spectra}
\label{optical_spectra}

For four of the eight extreme \nustar sources studied here, the
optical spectra were obtained from our dedicated followup program
with Keck (for J1444 and J1653; using the LRIS instrument), Magellan
(J0823; using the IMACS instrument), and the NTT (J1512; using the
EFOSC2 instrument).\footnote{Magellan program ID: CN2015A-87. NTT
  program ID: 093.B-0881.} Details of the observing runs and followup
campaign are provided in \citet{Lansbury17}.
For two sources (J1506 and J1534) the optical spectra are from the SDSS. 
For the remaining two sources (J0505 and J1410) the spectroscopic redshifts
and spectra are from the 6dF survey (\citealt{Jones04,Jones09}) and the Anglo-Australian
Telescope (AAT) observations of \citet{RadburnSmith06}, respectively.
The optical spectra are provided in the Appendix. 
The spectroscopic redshifts (see Table \ref{nustar_basic_table}) are
all robust, having been determined using $4$--$15$ detected
emission/absorption lines for each source (median of $9$ detected
lines per source), 
except in the case of J1444 where the redshift solution is based on two weakly
detected emission lines (most likely \civ and \ciii at
$z=1.539$).

All of the optical spectra show narrow emission lines and have
continua which appear consistent with being
dominated by the host galaxy. In five cases (J0505, J1410, J1506,
J1534, and J1653) the latter is confirmed by the identification of galactic
absorption lines.
These optical properties are congruous with the interpretation of these AGNs
as obscured systems, in agreement with the X-ray constraints.
To quantify the emission line properties, we fit the optical spectra for the major
  lines at rest-frame $3500$--$7000$\AA\ (e.g. \oii,
  \hbeta, \oiii, \oi, \halpha, \nii, and \sii)
  with the $\mathtt{pyspeckit}$ software following \citet{Berney15}
  and the general procedure in Koss et al.\ (2017, submitted).  We correct
  the narrow line ratios (\halpha/\hbeta) assuming an intrinsic
  ratio of $3.1$ and the \citet{Cardelli89} reddening curve.

For six sources with significantly detected \halpha emission
  lines (signal-to-noise of $S/N\gtrsim 4$; J0505, J0823, J1410, J1506, J1512, and
  J1534), the \halpha full widths at half maximum (FWHM) range from $269$ to
$538$~$\mathrm{km~s^{-1}}$, before correction for instrument
resolution. In no case is a second (broad-line) component
required to provide a statistically acceptable fit to the data. These results
confirm the visual classifications of these sources as narrow-line systems
\citep{Lansbury17}. We note that J1653 has only a weak detection
of \halpha, and J1444 is at high redshift ($z=1.539$) such that the above emission
lines are not in the redshifted spectrum.

\begin{figure}
\centering
\includegraphics[width=0.47\textwidth]{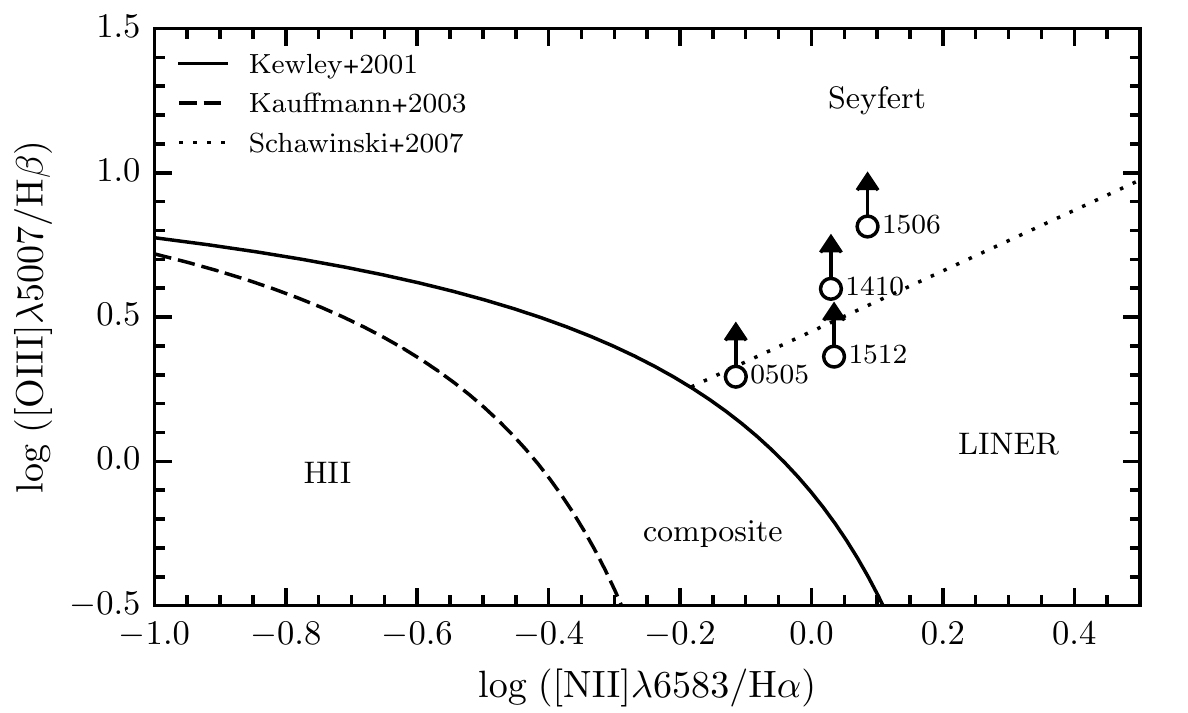}
\caption{Emission line ratios for the four sources where BPT
    diagnostics are possible. The solid line shows a theoretical
    maximum for starbursts (\citealt{Kewley01}), the dashed line shows an empirical
  threshold to separate star-forming \mbox{\ion{H}{2}} regions from AGNs
  (\citealt{Kauffmann03}), and the dotted line shows an empirical
  threshold to distinguish between Seyfert AGNs and LINER classifications (\citealt{Schawinski07}).
}
\label{bpt}
\end{figure}

For four sources (J0505, J1410, J1506, and J1512), it is
  possible to apply AGN emission-line
  diagnostics (e.g., \citealt{Kewley06}; \citealt{Veilleux87}) using the
  \nii/\halpha and \oiii/\hbeta emission-line flux ratio
  constraints. This is not possible for J0823 due to a
  gap in the spectrum, and for J1534 and J1653 due to the low $S/N$ of
  the key emission lines.
  Figure \ref{bpt} shows the location of the former four sources on
  the Baldwin-Phillips-Terlevich (BPT)
  diagram. All four sources fall into the AGN region based on the upper
  limits for the \hbeta line, which is weak to
  undetected ($S/N<3$). The weak \hbeta line emission is likely 
  due to extinction by dusty gas and has previously been observed for X-ray selected obscured AGNs,
  particularly in mergers (e.g. \citealt{Koss16a,Koss16b}).  
  We also note that \hbeta is undetected for J0823, J1534,
  and J1653, and even \oiii is undetected in the case of J1534.
  The seven $z<0.4$ extreme \nustar AGNs would thus be unidentified in
  any optical surveys requiring the detection of \hbeta.

\subsection{Host galaxies}
\label{host_galaxies}

\begin{figure*}
\centering
\includegraphics[width=1.0\textwidth]{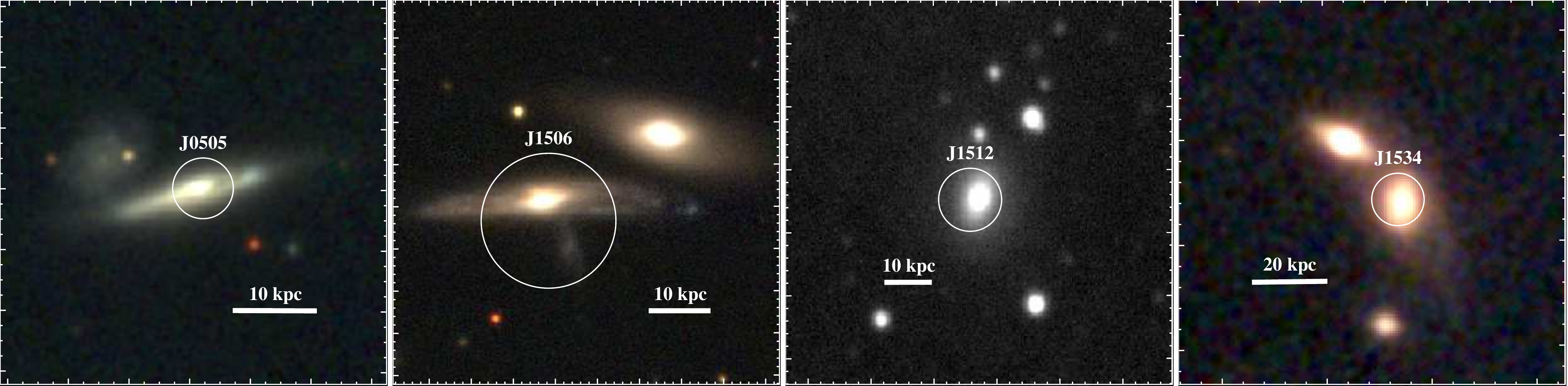}
\caption{Optical images for the extreme \nustar sources which have
  both a high CT likelihood and a well resolved host galaxy in the optical
  imaging. For J0505 (first panel; $z=0.036$), J1506 (second panel; $z=0.034$), and J1534 (fourth panel; $z=0.160$) we use
  Pan-STARRS ($g$, $r$, and $i$ band) color
  composites. For J1512 (third panel;
  $z=0.069$) we use NTT $R$-band imaging from our followup program.
  The white circles mark the X-ray positions: for J1506 we show the \nustar positional error circle ($16\arcsec$
  radius) while for J0505, J1512, and J1534 the circles mark the \xmm,
  \swiftxrt, and \chandra positions, respectively ($5\arcsec$,
  $5\arcsec$, and $2.5\arcsec$ radii shown, respectively). North is up and
  east is to the left. The major tickmarks indicate
    $10\arcsec$ offsets in R.A.\ (horizontal axes) and decl.\
    (vertical axes).
  Two of these \nustar-identified likely-CT AGNs (J1506 and
  J1534) belong to major mergers, with likely tidal features visible
  in both cases.}
\label{ds9_cutouts} 
\end{figure*}

The five lower redshift ($z<0.2$) extreme \nustar sources (J0505, J1410, J1506, J1512, and
J1534) have well resolved host galaxies at optical
wavelengths, while the higher redshift sources are consistent with point-source emission. 
Four of the five lower redshift sources are likely-CT systems based on our X-ray analyses, and
also have relatively high quality optical coverage from Pan-STARRS
(PS1; \citealt{Chambers16}) or
our own ESO-NTT imaging (see Figure \ref{ds9_cutouts}). The other lower redshift source
(J1410), on the other hand, is Compton-thin, and is limited to
low-quality optical coverage from photographic plate observations. 
Here we comment on the host galaxies, and nearby companion galaxies,
for the lower redshift sources. 

\paragraph{J0505} The optical counterpart is 2MFGC~04170, a highly inclined
disk galaxy. The Pan-STARRS coverage of 2MFGC~04170 reveals spatially extended
emission at $\approx 12''$ offset (or a projected separation of
$\approx 9$~kpc), and at a position angle of $\approx 70$\degrees,
which appears consistent with being a companion galaxy to 2MFGC~04170
(see Figure \ref{ds9_cutouts}). We hereafter refer to this
second companion source as J050601.2--235002.6. 
Since this source had no available redshift information,
we performed followup spectroscopy with Keck (provided in the Appendix).
We find that J050601.2--235002.6 lies at $z=0.137$, and is therefore a background
galaxy which is coincidentally aligned along the line-of-sight, rather
than being a merging companion to 2MFGC~04170.

\paragraph{J1506} The optical counterpart is UGC~09710, an edge-on Sb spiral galaxy
belonging to a close spiral-spiral galaxy pair in an early-stage major
merger (see Figure \ref{ds9_cutouts}), and separated from 
its similar mass partner galaxy (IC~1087; $z=0.035$; S0-a type) by
$\approx 16$~kpc in projection \citep{Yuan12}. 
Physical disturbances resulting from the major merger could
potentially be related to an increase in the central gas content.
In the Appendix we present a Palomar optical spectrum
for the companion galaxy (IC~1087), which shows a possible AGN (also
consistent with a LINER classification) with a dominant galaxy continuum. 
\oiii and \hbeta are undetected for the companion galaxy
(presumably due to host-galaxy dilution), and the \nii:\halpha line
strength ratio is very high, 
but is likely affected by stellar absorption. 
For this companion galaxy, there is no additional evidence from the
\wise colors for an AGN, and the source is undetected in the current X-ray coverage.

\paragraph{J1410} The available photographic
plate coverage (from the UK Schmidt Telescope) shows an extended host
galaxy, but the low data quality preclude type and disturbance classifications. 
Nevertheless, there do not appear to be any nearby (massive) companion galaxies.

\paragraph{J1512} We have obtained $R$-band imaging with the ESO-NTT
(shown in Figure \ref{ds9_cutouts}), which is in visual agreement with the host being a relatively undisturbed early
type galaxy. The neighbouring optical sources are consistent with
being unresolved point sources, with FWHMs similar to the seeing
($\approx 1.5''$), and are therefore unlikely to be associated with J1512.

\paragraph{J1534} The Pan-STARRS imaging (Figure~\ref{ds9_cutouts})
shows good evidence that the
optical host galaxy (SDSS~ J153445.80+233121.2; $z = 0.160$) is
undergoing a major merger with a narrowly
offset companion galaxy (SDSS~J153446.19+233127.1; no
spec-$z$); the respective galaxy nuclei are separated by
$\approx 8''$ (or $\approx 22$~kpc in projection), and likely extended tidal features
are visible. The merger stage is not clear. We present Palomar spectroscopic
followup for the companion galaxy in the Appendix, although there are
no significantly detected emission or absorption features.

\ \ \ 

A notable feature of the galaxies is that both
  J0505 and J1506 have close to edge-on geometries, which could
  contribute at least some of the observed X-ray obscuration.
  The axis ratios of the host galaxies are $b/a=0.24$ and $0.23$ for
  J0505 and J1506, respectively, based on isophotal fitting of the galaxy
  images in Figure \ref{ds9_cutouts} (using the IRAF task
  $\mathtt{ellipse}$). The remaining two likely-CT sources (J1512 and
  J1534), on the other hand, have axis ratios exceeding $b/a=0.6$.
  Although the source numbers are currently small, the above implies a
  relatively high fraction ($50\pm 33 \%$) of close to edge-on systems for CT AGNs
  selected by \nustar. 
  For comparison, only $\approx 16\%$ of the
  general hard X-ray selected AGN population have $b/a<0.3$, based 
  on isophotal analyses for the \swiftbat AGN sample
  (\citealt{Koss11}). Although the difference is only weakly
  significant, a similar result has also been reported for CT AGNs
  selected with \swiftbat (\citealt{Koss16a}). 
  Other studies, however, find that edge-on galaxy inclinations are not
  clearly related to CT absorption (e.g., \citealt{Annuar17};
  \citealt{Buchner17}).

\subsubsection{A high fraction of galaxy mergers for the Compton-thick
AGNs?}

It is interesting that two of the four likely-CT AGNs 
(J0505, J1506, J1512, and J1534) are hosted by galaxy major mergers
(see Figure \ref{ds9_cutouts}).
To assess the statistical significance of the apparently high merger
fraction for these
extreme \nustar serendipitous survey AGNs ($f_{\mathrm{merger}} =50\pm 33\%$; % 50.0\pm 31.7\%$
the errors represent binomial uncertainties), we can search for
similar merging systems in the sample of non-extreme (or
``normal'') serendipitous survey AGNs. To this end, from the overall
serendipitous survey sample, we apply a cut of
$\mathrm{BR}_{\mathrm{Nu}}<1.7$, thus limiting to those sources which do
not have very hard \nustar spectra (based on the \brnu threshold in Section
\ref{selection}). We limit this comparison sample to source redshifts of
$0.01<z<0.2$, thus matching the redshift range of the
four extreme sources. 
We exclude two sources from the sample which are
likely strongly associated with the science targets of their \nustar
observations (similar to the exclusion of J2028 from the extreme
sample; see Section \ref{selection}).
These cuts leave $36$ normal \nustar sources. Finally, we limit the sample to the $26$ (out of $36$) sources
which are covered by Pan-STARRS observations, and
therefore have optical coverage which is of comparable quality to the
four extreme \nustar sources. As a result, the comparison of visual
merger classifications between the two different samples is
unlikely to be significantly affected by variations in optical imaging
sensitivity. 
The comparison sample is matched in
X-ray luminosity distribution to the extreme \nustar AGNs (with a
Kolmogorov-Smirnov test p-value of $0.8$). 

\begin{figure}
\centering
\includegraphics[width=0.47\textwidth]{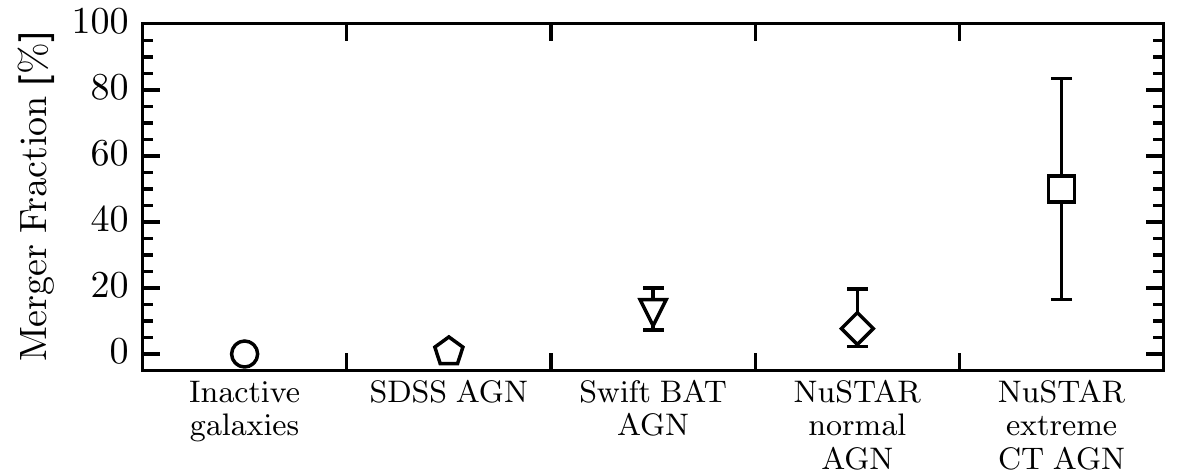}
\caption{The fraction of host galaxies in major
  mergers, for \nustar serendipitous survey sources at $z<0.2$.
The fraction is shown for two subsets of the serendipitous survey: (1)
the extreme AGNs (square) with very hard X-ray spectra and evidence for
CT obscuration (J0505, J1506, J1512, and J1534; i.e., those
discussed in this work) and (2) ``normal'' \nustar
AGNs (diamond).
We also compare to the 
major-merger fraction for \swiftbat AGNs (triangle; \citealt{Koss10}), and for
inactive galaxies and SDSS AGNs matched to the \swiftbat sample
(circle and pentagon, respectively; \citealt{Koss10}; the error bars
are smaller than the data points). Uncertainties are shown at the $90\%$
confidence level.
}
\label{merger_fraction}
\end{figure}

Of the $26$ normal AGNs, we identify one which has 
evidence for a galaxy major merger, with a comparably sized 
companion galaxy lying at the same redshift and offset by a projected distance
of $\approx 25$~kpc. There are an additional two normal AGNs with possible evidence for
mergers, although the candidate companion galaxies are relatively small in
size, with unknown redshifts. We conservatively assume that two of the
normal AGNs are in major mergers with $<30$~kpc-separation companions. 
Our estimate for the major-merger fraction of normal \nustar
AGNs is therefore $f_{\mathrm{merger}}=8^{+12}_{-5}\%$. 
This is in agreement with the ($<30$~kpc-separation) major-merger fraction for
\swiftbat AGNs ($f_{\mathrm{merger}}=13^{+7}_{-5}\%$; 
\citealt{Koss10}). 
Figure \ref{merger_fraction} compares the above merger fractions. 
We additionally compare with low redshift inactive galaxies and
optical Type~2 AGNs (both from the SDSS), which
are matched to the \swiftbat sample (\citealt{Koss10}), and have
very low merger fractions compared to the \swiftbat and extreme \nustar AGNs.
At low significance levels of $1.8\sigma$ and $1.7\sigma$ (according to the Fisher exact
probability test), 
the extreme (very hard, CT) \nustar AGNs have a higher merger
fraction than both the normal \nustar AGNs and the \swiftbat AGNs, respectively. 
This could be a result of Compton-thick phases of black hole growth 
being more strongly linked (than less-obscured phases) to the
merger stage of the galaxy evolutionary sequence.

The above result is of interest given recent findings
for other AGN samples. 
\citet{Kocevski15} find evidence that highly obscured
($N_{\rm H}\gtrsim 3\times 10^{23}$~\nhunit) AGNs at $z\sim1$ have a
higher frequency of merger/interaction morphologies relative to less
obscured AGNs matched in redshift and luminosity. 
Furthermore, \citet{Koss16a} noted a high 
close ($<10$~kpc) merger fraction for likely-CT \swiftbat
AGNs at $z\lesssim 0.03$ ($f_{\mathrm{merger}}=22\%$; i.e., $2/9$).
The recent study of \citet{Ricci17} indicates 
a possible connection between the late stages
of galaxy mergers and high AGN obscuration, in a sample of local
luminous and ultra-luminous infrared galaxies (U/LIRGs), using a
combination of dedicated and archival X-ray observations.  
Taken together, the results may suggest a departure
from simple orientation-based unified models of AGN obscuration, and
indicate an evolutionary scenario where highly obscured phases of
black hole growth can be associated with a merger-driven
increase in the circumnuclear gas content (e.g.,
\citealt{Sanders88,Draper10,Treister10a}). 
An increased sample size and deeper imaging would help to
further test the CT AGN-merger connection using the \nustar serendipitous
survey.

%%%%%%%%%%%%%%%%%%%%%%%%%%%%%%%%%%%%%%%%%%%%%%%%%%%%%%%%%%%%%%%%%%%%%%
\section{The prevalence of Compton-thick absorption}
\label{prevalence}
%%%%%%%%%%%%%%%%%%%%%%%%%%%%%%%%%%%%%%%%%%%%%%%%%%%%%%%%%%%%%%%%%%%%%%

\begin{figure}
\centering
\includegraphics[width=0.47\textwidth]{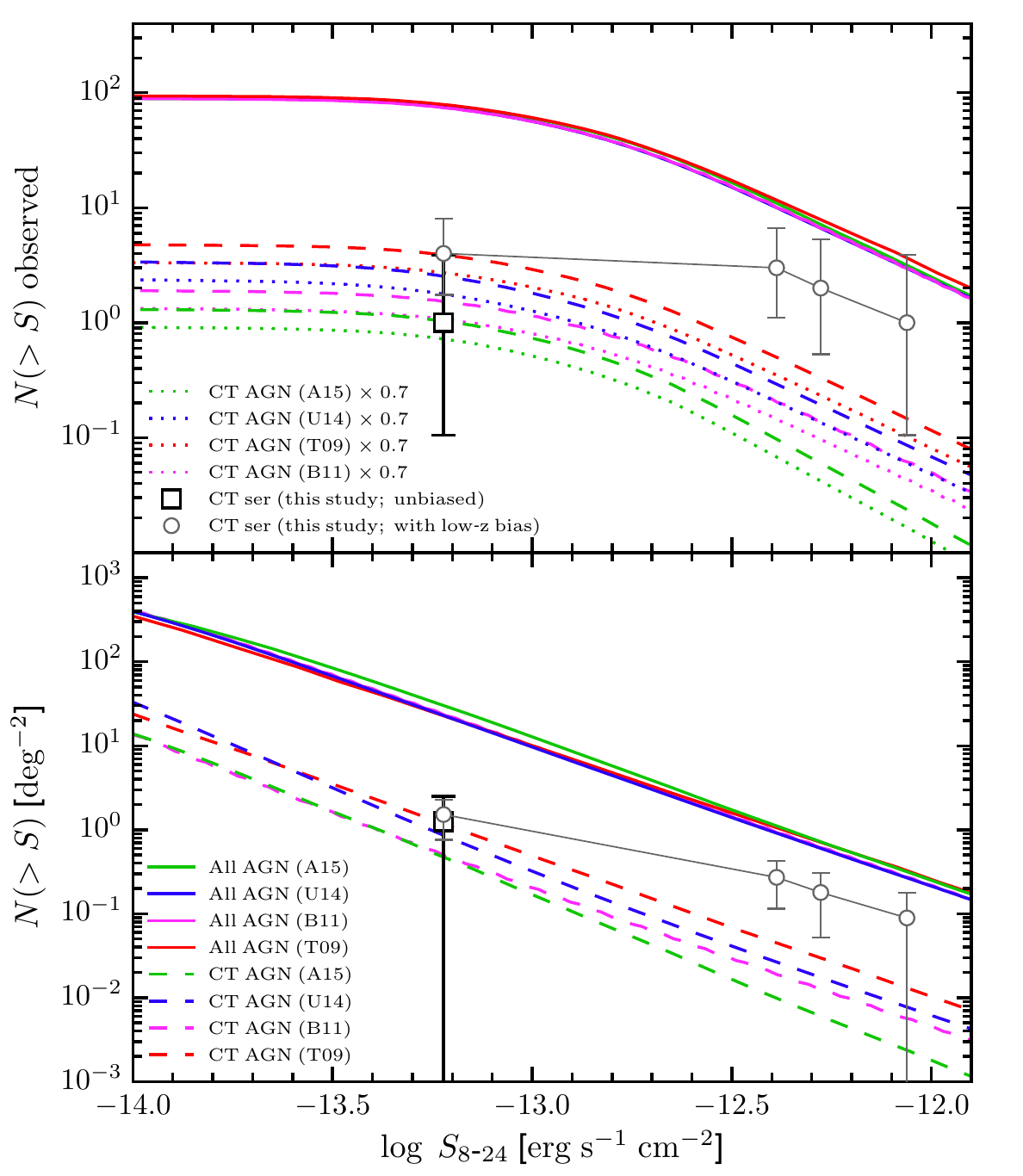}
\caption{Top panel: observed cumulative number counts (and $90\%$
  CL uncertainties), as a function
  of $8$--$24$~keV flux ($S_{\rm 8\mbox{-}24}$), for the CT AGNs identified in the
  \nustar serendipitous survey. 
  The gray circles show the number counts for all four CT AGNs.
  The black square shows the modified number counts when removing the
  three low-redshift CT AGNs (J0505, J1506, and J1512; see Section \ref{prevalence}).
  We compare to predicted tracks for CT
  AGNs (dashed lines) and all AGNs (solid lines) based
  on the models of A15, U14, B11, and T09. The
  dotted lines show modifications of the CT model tracks to account
  for the spectroscopic incompleteness of the serendipitous
  survey. Lower panel: ``intrinsic'' cumulative number density (and $68\%$
  CL uncertainties) as a function of flux. 
}
\label{number_counts} 
\end{figure}

We have taken advantage of the relatively large sample size of the \nustar
serendipitous survey to identify rare highly obscured AGNs. 
While all of the eight extreme sources investigated are 
consistent with being highly obscured, four in particular are likely
CT (J0505, J1506, J1512, and J1534). A fifth source (J1653) is 
a CT candidate based on the X-ray analysis, but this result is in
tension with the indirect constraints (see Section \ref{indirect}).
Here we assess how the observed number of CT AGNs in the \nustar
serendipitous survey compares with the number expected from AGN
population models, which are informed by the
results from previous (primarily $<10$~keV) X-ray surveys.
We consider the hard band ($8$--$24$~keV) selected serendipitous survey sample, since this is the
energy band in which \nustar is uniquely sensitive, and Galactic
latitudes of $|b|>10$\degrees (i.e., out of the Galactic plane). We
conservatively exclude J1653. 
The top panel of Figure \ref{number_counts} shows the observed
(cumulative) number of CT sources as a function of limiting flux, 
and these results are compared to model predictions for the observed numbers of CT AGNs
and all AGNs. For these predictions, we fold the area-sensitivity curve of the
serendipitous survey through 
models for the evolution of the X-ray luminosity function (XLF) and the \nh
distribution of AGNs, 
from \citet[][hereafter T09]{Treister09}, \citet[][hereafter U14]{Ueda14},
\citet[][hereafter A15]{Aird15a}, and the updated version of
\citet[][hereafter B11]{Ballantyne11}. The updates to the B11 model
are summarised in \citet{Harrison16}. 
We additionally show, in the 
lower panel of Figure \ref{number_counts}, the ``intrinsic'' cumulative number densities
[i.e., the sky number counts before accounting for the survey sensitivity; $N(>S)$, in units of
$\mathrm{deg}^{-2}$].

In Figure \ref{number_counts} the gray circle data points show the number counts
for all four CT AGNs.
There is an apparent excess in the CT number counts at high fluxes,
compared to the model predictions. This excess may be expected given that the
three lowest-redshift, highest-flux sources (J0505,
J1506, and J1512; $z<0.07$) show evidence for being weakly associated with 
the \swiftbat AGN targets of the \nustar observations (see Section
\ref{associated_note}), and also given that galaxy
clustering tends to be high around BAT AGNs (e.g., \citealt{Koss10,Cappelluti10}). 
In Figure \ref{number_counts} we also show the CT 
number counts using J1534 only (i.e., excluding
J0505, J1506, and J1512; black square data point). Although not
particularly constraining, this brings the number counts into better
agreement with all of the models (T09, B11, U14, A15, and
\citealt{Gilli07}), suggesting consistency with a wide range of 
intrinsic CT fractions\footnote{The CT fraction is defined here as the fraction of all AGNs which are CT.}
ranging from $f_{\rm CT}\approx 10$--$40\%$, at least for $z>0.07$. 
For comparison, Zappacosta et al.\ (2017, submitted) study the
X-ray spectral properties of \nustar extragalactic survey sources and
find that the range of CT fractions allowed by their sample is broad
($f_{\rm CT} \approx 10$--$70\%$). The \nustar survey constraints on
$f_{\rm CT}$ are therefore in broad agreement with $z\gtrsim 0.1$ constraints from soft ($<10$~keV)
X-ray observatories ($f_{\rm CT} \approx 30$--$50\%$; e.g.,
\citealt{Brightman12,Brightman14,Buchner15}).

\begin{figure}
\centering
\includegraphics[width=0.47\textwidth]{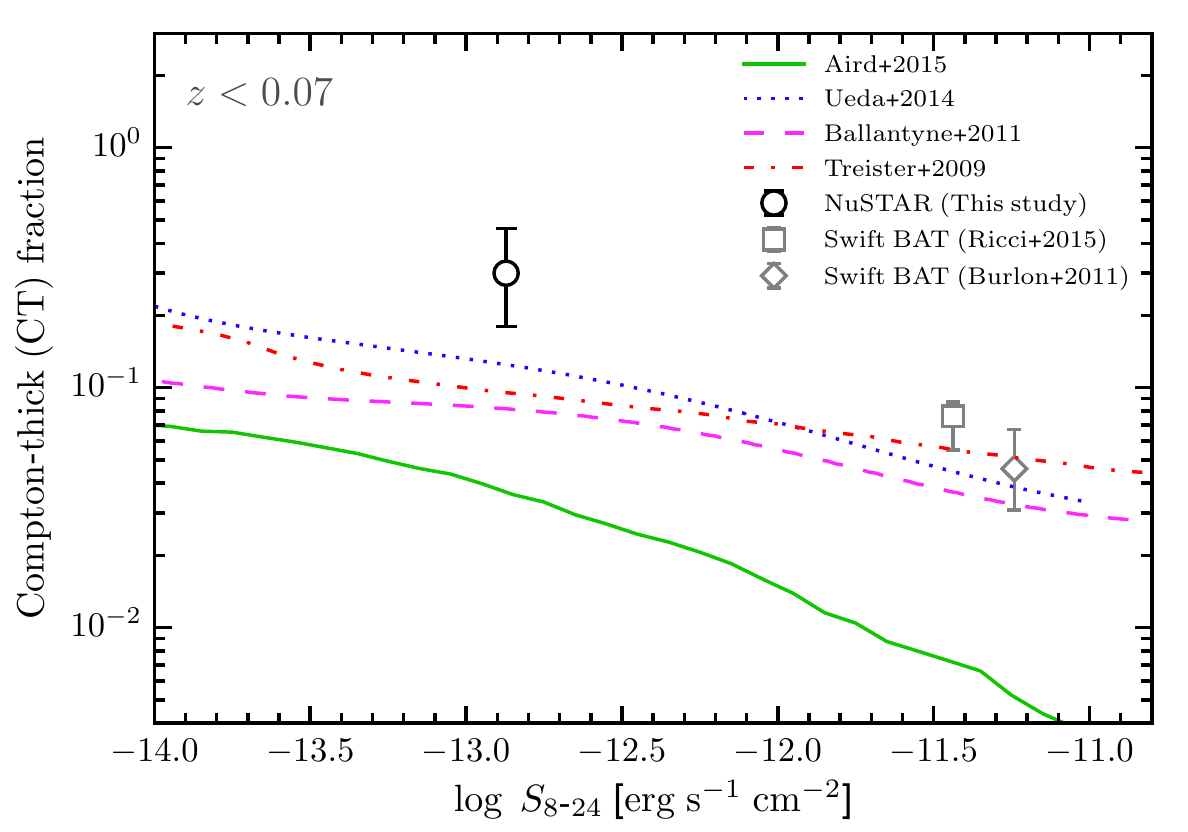}
\caption{Observed CT fraction (relative to all
  AGNs) as a function of $8$--$24$~keV flux limit, for $z<0.07$. The
  black circle data point shows the \nustar serendipitous survey
  constraint from this work. The grey data points
  show constraints using the three-year (diamond; \citealt{Burlon11}) and $70$-month
  (square; \citealt{Ricci15}) \swiftbat surveys, respectively. We
  compare with model predictions based on A15 (green solid line), U14
  (blue dotted line), B11 (pink dashed line), and T09 (red dash-dotted
  line).
}
\label{fCT_S} 
\end{figure}

However, it is important to consider independently the low-redshift
  ($z<0.07$) regime, where we have detected
  the highest numbers of CT AGNs. Although the overall number
  counts in this regime may have an upwards excess with respect to model predictions
  (as mentioned above), the CT {\it fraction} should be unaffected.
  The observed CT fraction for the $z<0.07$ \nustar serendipitous
  survey sample is $f_{\rm CT}^{\rm obs}=30^{+16}_{-12}\%$ ($68\%$ CL
  binomial uncertainties). The intrinsic X-ray luminosity range of this subsample is
  $41.3<\log ( L_{\rm 10\mbox{-}40keV} / \mathrm{erg\ s^{-1}} )
  <44.0$. Figure \ref{fCT_S} compares our observational constraint to 
  model predictions as a function of $8$--$24$~keV flux.
  We find a higher CT fraction than is expected from the models. The
  difference is statistically significant in one case ($>3\sigma$; comparing
  to A15) and at lower significance levels for the remaining 
  models ($<3\sigma$; comparing to T09, B11 and U14), 
  In Figure \ref{fCT_S} we additionally compare with data points for the
  higher-flux \swiftbat survey (\citealt{Burlon11,Ricci15}), for which we
  have converted to the $8$--$24$~keV \nustar band assuming
  $\Gamma_{\rm eff}=1.9$. At present, the origin of the
  high observed CT fraction at $z<0.07$ is unclear. A likely explanation is 
  that the current models are not well constrained for the new parameter space
  probed with \nustar, in which case the AGN population models require
  updating. An alternative possibility, however, is that $f_{\rm
    CT}^{\rm obs}$ is boosted
  due to a real connection between CT absorption and the large-scale environment,
  in combination with \nustar having preferentially targetted (at
  $z<0.07$) fields with relatively high galaxy densities (e.g.,
  fields around \swiftbat AGNs).

Finally, we note that the number of CT AGNs presented here could be
a lower limit to the total number within the \nustar serendipitous survey as there are additional sources,
not included in this work, which have band-ratio limits consistent
with a large range in column density (e.g., see Figure \ref{br_z}), and any
CT sources with relatively soft spectral shapes could potentially be
missed by our initial selection (Section \ref{selection}). 
Alternative approaches (e.g., detailed X-ray or multi-wavelength
analyses of the broader sample) may tease out additional CT AGNs
within the sample.
However, large improvements on the constraints presented here will require
further survey data from sensitive hard X-ray
missions. Further data will be provided by the continued \nustar operations,
which are likely to increase the serendipitous sample to $\gtrsim 1000$
sources, and potentially by future high-sensitivity $>10$~keV observatories (e.g., the
High-Energy X-ray Probe, or {\it HEX-P}, mission concept currently
under study; PI F.~Harrison; see \citealt{Brandt15} for a brief overview).

%%%%%%%%%%%%%%%%%%%%%%%%%%%%%%%%%%%%%%%%%%%%%%%%%%%%%%%%%%%%%%%%%%%%%%
\section{Summary}
\label{summary}
%%%%%%%%%%%%%%%%%%%%%%%%%%%%%%%%%%%%%%%%%%%%%%%%%%%%%%%%%%%%%%%%%%%%%%

In this paper we have searched for the most extreme sources in the
\nustar serendipitous survey, in terms of having very hard
spectral slopes ($\mathrm{BR}_{\rm Nu}\ge 1.7$). The 
eight selected sources are all candidates for being highly obscured
AGNs. A detailed look at the broad-band ($0.5$--$24$~keV) X-ray data
available, and the multiwavelength properties of these sources, has
yielded the following main results:

\begin{itemize}

\item The X-ray spectral analyses find that three of the extreme \nustar sources
  (J0505, J1506, and J1512) are newly identified robust Compton-thick (CT) AGNs at low redshift ($z<0.1$).
An additional source at higher redshift (J1534) is likely CT. The
remaining four extreme sources are consistent with being CT or at least
moderately absorbed; see Section \ref{xray_results}. 

\item Most (three out of four) of the likely-CT AGNs identified with \nustar would not have been identified as
  highly obscured systems based on the low energy ($<10$~keV) X-ray
  coverage alone.
J1506 is a notable example: a newly uncovered 
  CT AGN in the nearby universe ($z=0.034$; $N_{\rm H}>
  2\times 10^{24}$~\nhunit; $L_{\rm X}\approx 2\times
  10^{43}$~\ergpersec), hosted by a previously known galaxy
  major merger; see Sections \ref{xray_results} and \ref{host_galaxies}.

\item For all eight extreme sources, 
  the optical spectra show evidence for narrow line AGNs
  or galaxy-dominated spectra, supporting the X-ray
  classifications as obscured and CT AGNs; see Section
  \ref{optical_spectra}.
  Measurements of the X-ray to MIR luminosity ratio, an indirect
  absorption diagnostic, are also broadly congruent with the X-ray
  classifications. Two sources (J1506 and J1534) have particularly extreme
  ratios, lying even lower in $L_{\rm X}/L_{\rm MIR}$ than the
  well-known CT AGNs in the local Universe; see Section
  \ref{indirect}.

\item A high fraction ($50\pm 33\%$) of the likely-CT AGNs are hosted by galaxy
  major mergers. This is higher than the major-merger fractions for
  ``normal'' \nustar serendipitous survey sources
  and for \swiftbat AGNs, at a low significance level, motivating
  larger future studies; see Section \ref{host_galaxies}

\item We estimate the number counts of CT AGNs for the hard band
  ($8$--$24$~keV) selected serendipitous survey sample
  at $|b|>10$\degrees. The number counts are broadly
    harmonious with AGN population
  models over the main redshift range of the survey ($0.1 \lesssim
  z\lesssim 2$), but there is disagreement at low redshifts
  ($z<0.07$) where we find evidence for a
  high observed CT fraction of $f_{\rm CT}^{\rm obs}=30^{+16}_{-12}\%$; see
  Section \ref{prevalence}.

\end{itemize}

%%%%%%%%%%%%%%%%%%%%% acknowledgements %%%%%%%%%%%%%%%%%%%%%%%%%%%%%%%
\section*{Acknowledgements}

The authors first thank the anonymous referee for the positive and constructive review.
We acknowledge support from: 
a Herchel Smith Postdoctoral Research Fellowship of the University of Cambridge (G.B.L.);
the Science and Technology
Facilities Council (STFC) grants ST/I001573/1 (D.M.A.) and ST/J003697/2 (P.G.);
the ERC Advanced Grant FEEDBACK 340442 at the University of Cambridge (J.A.); 
the NASA Earth and Space Science Fellowship Program, grant NNX14AQ07H (M.B.);
CONICYT-Chile grants FONDECYT Regular 1141218 (F.E.B.), FONDECYT 1120061 and 1160999 (E.T.), and Anillo ACT1101 (F.E.B. and E.T.); 
the Center of Excellence in Astrophysics and Associated Technologies (PFB 06; F.E.B. and E.T.);
the Ministry of Economy, Development, and Tourism's Millennium Science Initiative through grant IC120009, awarded to The Millennium Institute of Astrophysics, MAS (F.E.B.);
ASI/INAF contract I/037/12/0-011/13 (A.C., A.M., and L.Z.); 
and {\em Chandra} grants GO5-16154X and GO6-17135X (J.A.T.).
We thank Yoshihiro Ueda and Roberto Gilli for providing number counts predictions.
This work was supported under NASA Contract No.\ NNG08FD60C, and made use of data from the \nustar mission, a project led by the California Institute of Technology, managed by the Jet Propulsion Laboratory, and funded by the National Aeronautics and Space Administration. We thank the \nustar Operations, Software and Calibration teams for support with the execution and analysis of these observations. This research has made use of the \nustar Data Analysis Software (NuSTARDAS) jointly developed by the ASI Science Data Center (ASDC, Italy) and the California Institute of Technology (USA).

Facilities: \chandra, ESO La Silla, Keck, Magellan, \nustar, Palomar, Pan-STARRS, SDSS, {\it Swift}, {\it WISE}, \xmm.
%%%%%%%%%%%%%%%%%%%%%%%%%%%%%%%%%%%%%%%%%%%%%%%%%%%%%%%%%%%%%%%%%%%%%%

\bibliography{bibliography.bib}{}

\appendix
\label{appendix}

\section{A.1 Optical spectra for the extremely hard \nustar
  serendipitous survey sources} 

\begin{figure*}
\centering
\begin{minipage}[l]{0.495\textwidth}
\includegraphics[width=\textwidth]{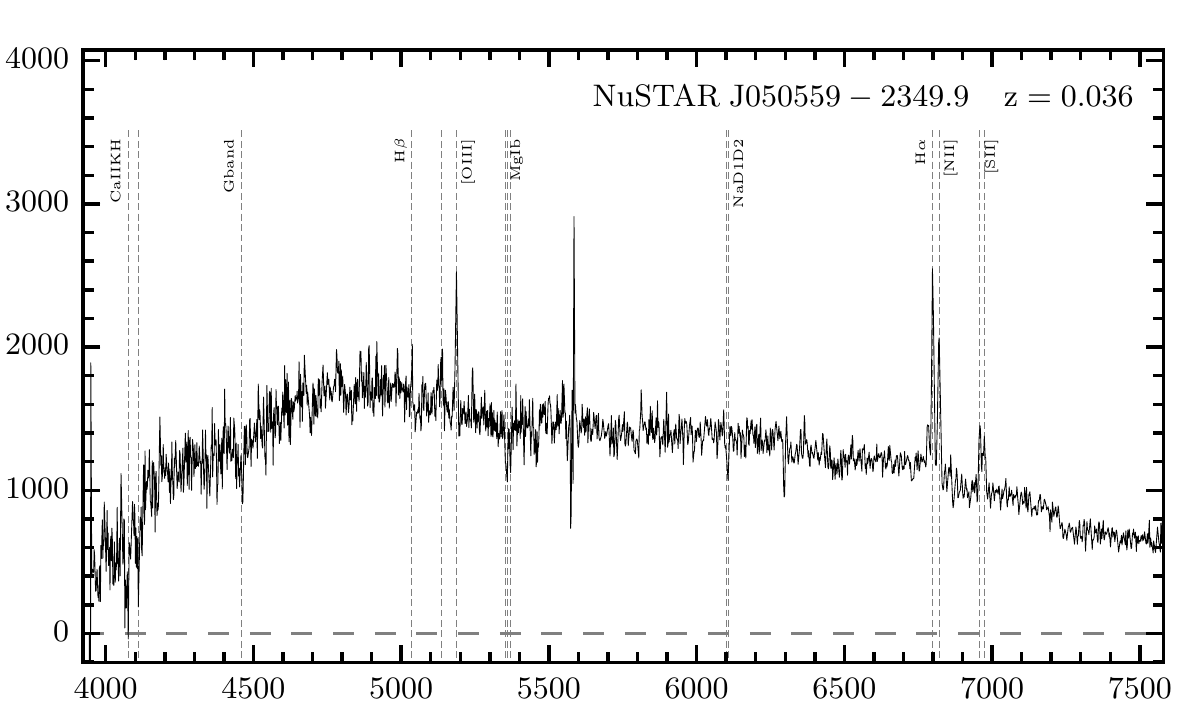}
\end{minipage}
\begin{minipage}[l]{0.495\textwidth}
\includegraphics[width=\textwidth]{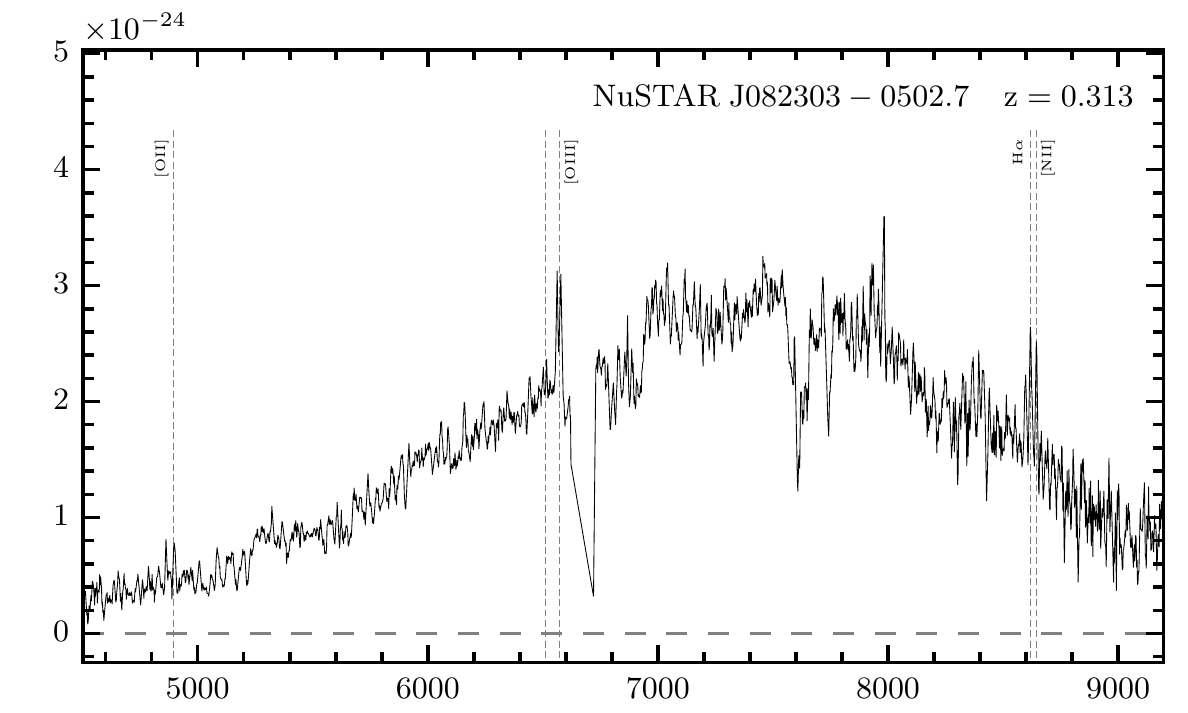}
\end{minipage}
\begin{minipage}[l]{0.495\textwidth}
\includegraphics[width=\textwidth]{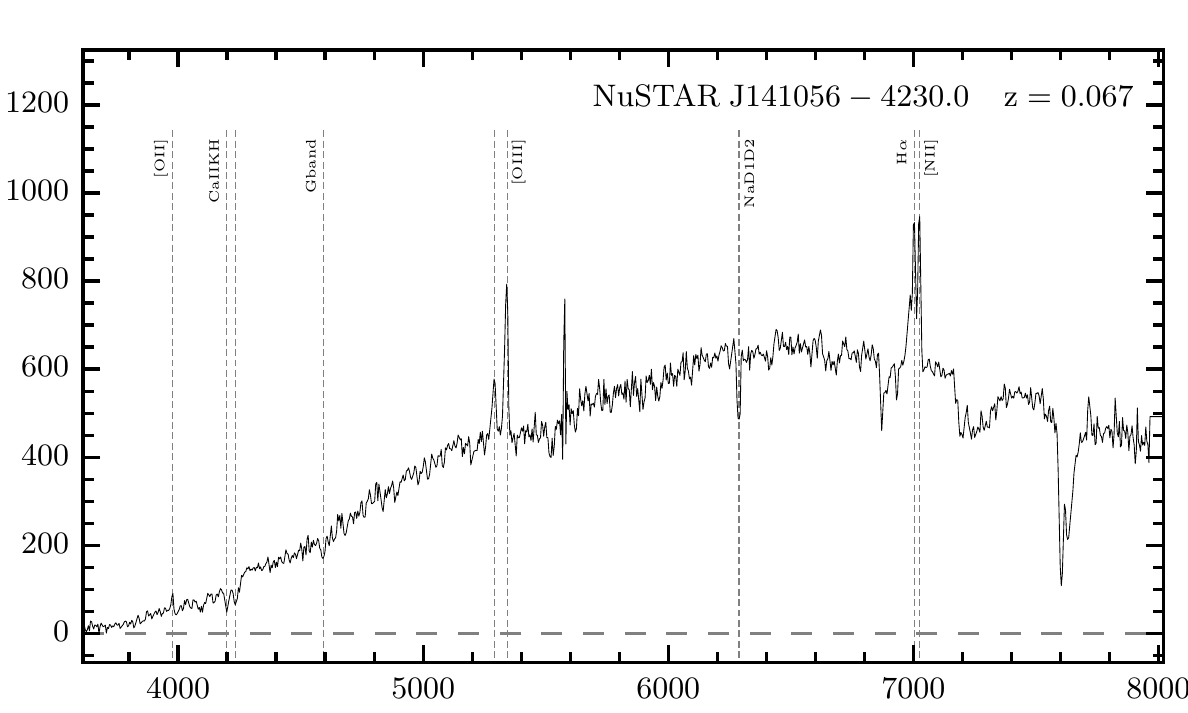}
\end{minipage}
\begin{minipage}[l]{0.495\textwidth}
\includegraphics[width=\textwidth]{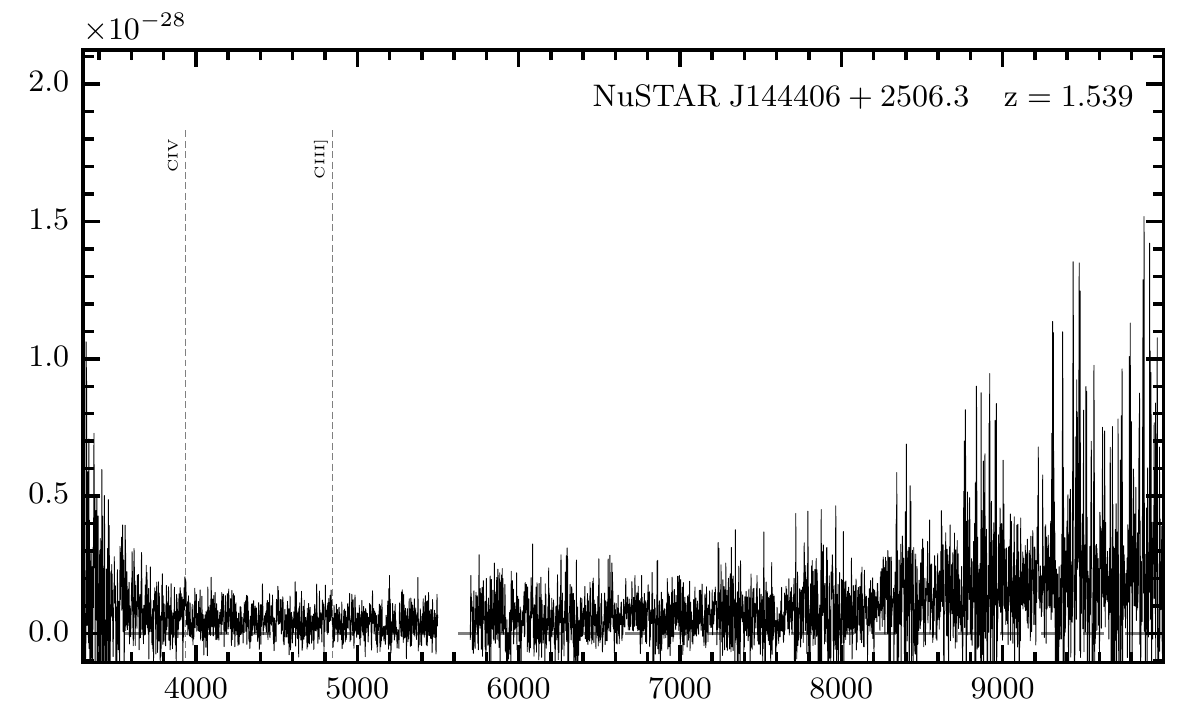}
\end{minipage}
\begin{minipage}[l]{0.495\textwidth}
\includegraphics[width=\textwidth]{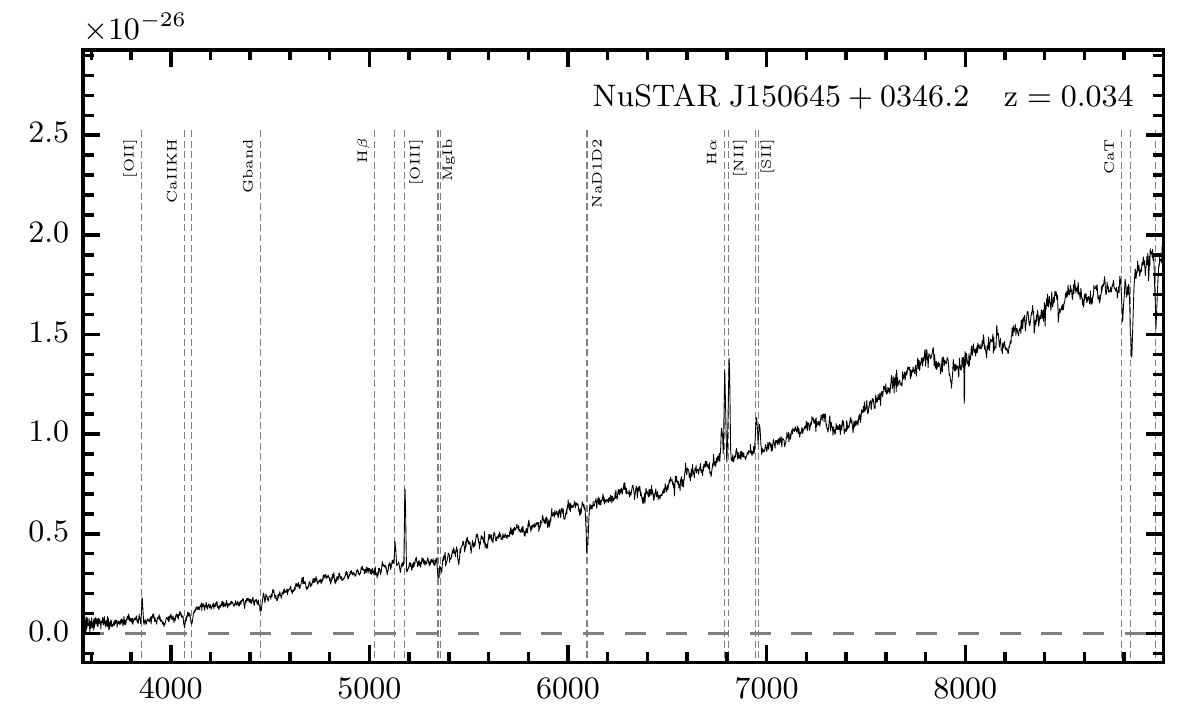}
\end{minipage}
\begin{minipage}[l]{0.495\textwidth}
\includegraphics[width=\textwidth]{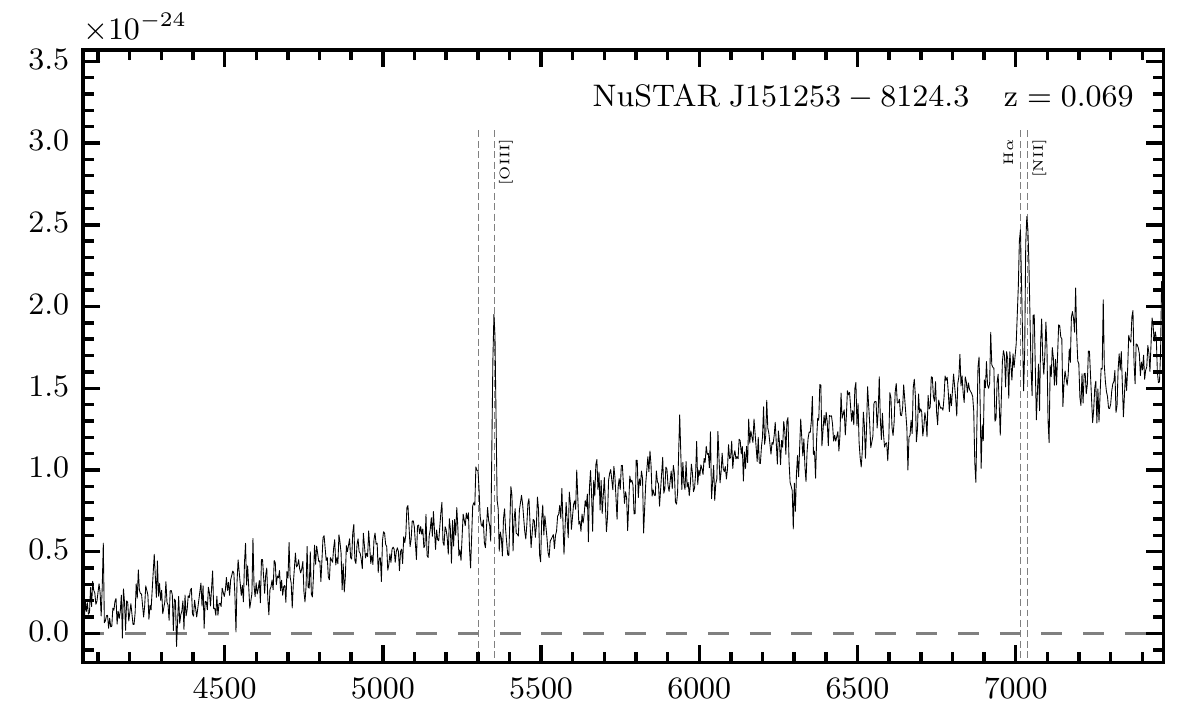}
\end{minipage}
\begin{minipage}[l]{0.495\textwidth}
\includegraphics[width=\textwidth]{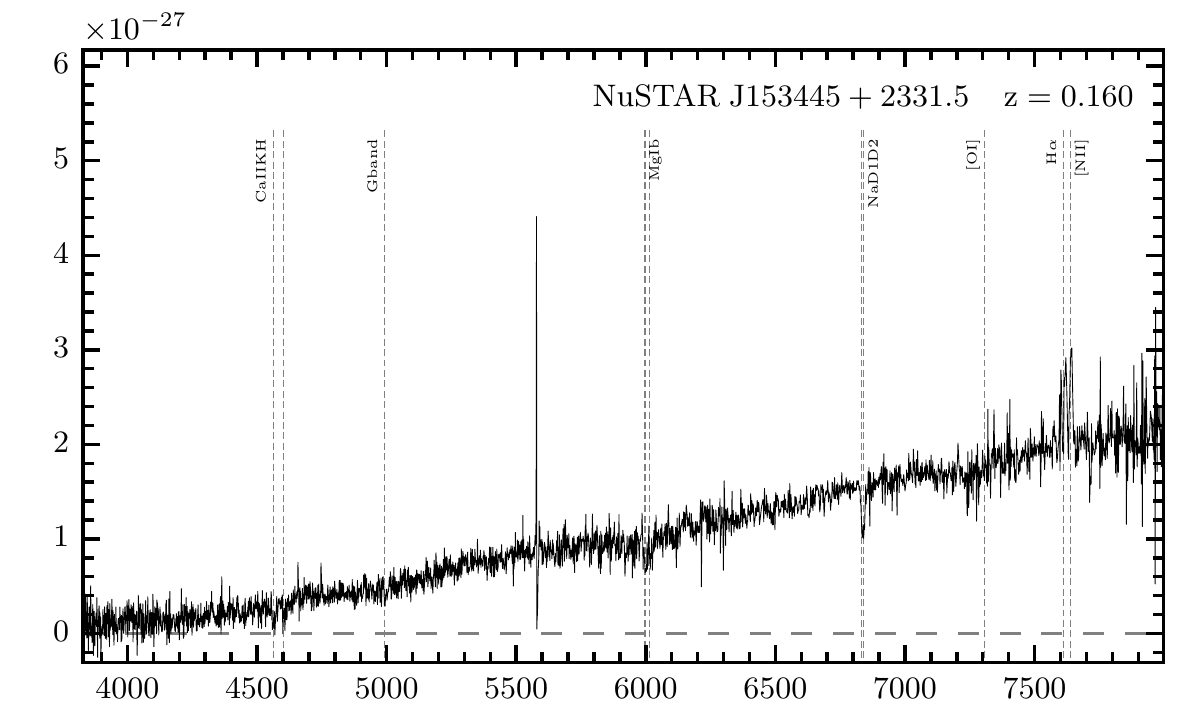}
\end{minipage}
\begin{minipage}[l]{0.495\textwidth}
\includegraphics[width=\textwidth]{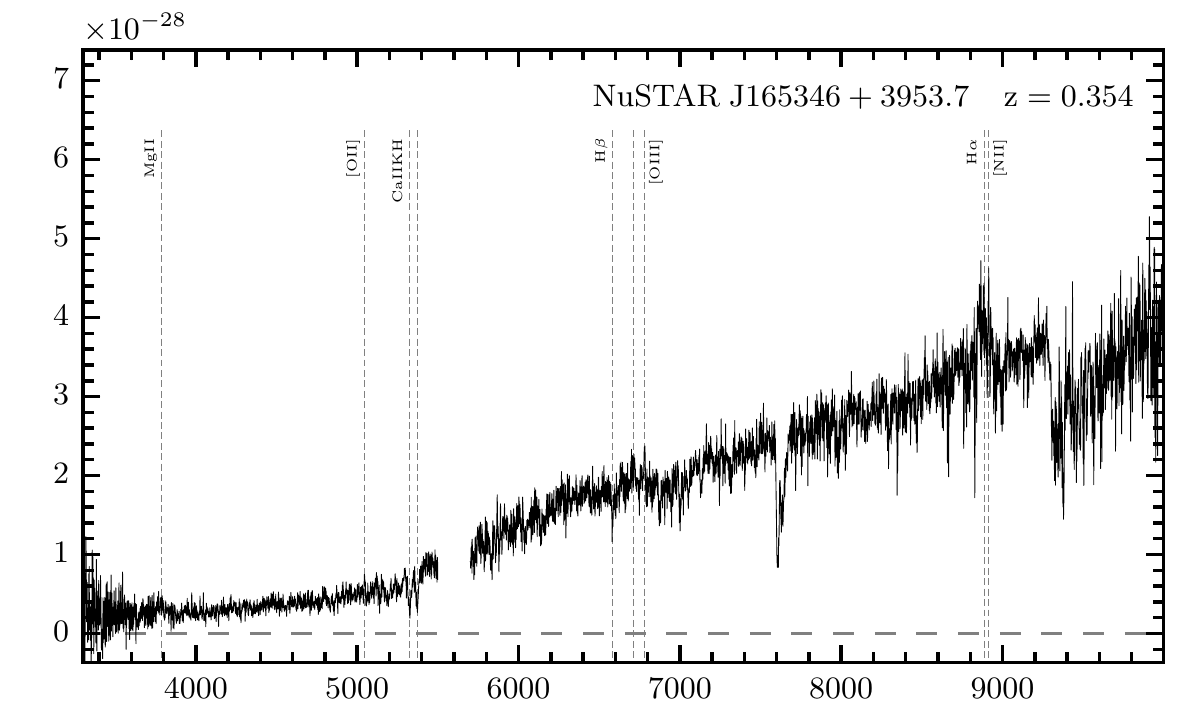}
\end{minipage}
\caption{Optical spectra for the extreme \nustar serendipitous survey
  sources. The horizontal axis shows the observed-frame wavelength in
  units of \AA. The vertical axis shows the flux ($f_{\mathrm{\nu}}$)
  in units of \fnuunit for all sources except J0505 and J1410, for
  which the vertical axis shows the counts. The vertical dashed gray lines mark
  the emission and absorption lines identified.}
\label{optspecfig}
\end{figure*}

  Here we provide the optical spectra (Figure \ref{optspecfig})
  for the eight extreme \nustar AGNs, which are discussed in
  Section \ref{optical_spectra}.
  The identified emission and absorption
  lines are highlighted in Figure \ref{optspecfig}, and are tabulated
  in Appendix A.2 of \citet{Lansbury17}.

\section{A.2 Optical spectra for companion galaxies}

\subsection{J0505}

\begin{figure}
\centering
\includegraphics[width=0.47\textwidth]{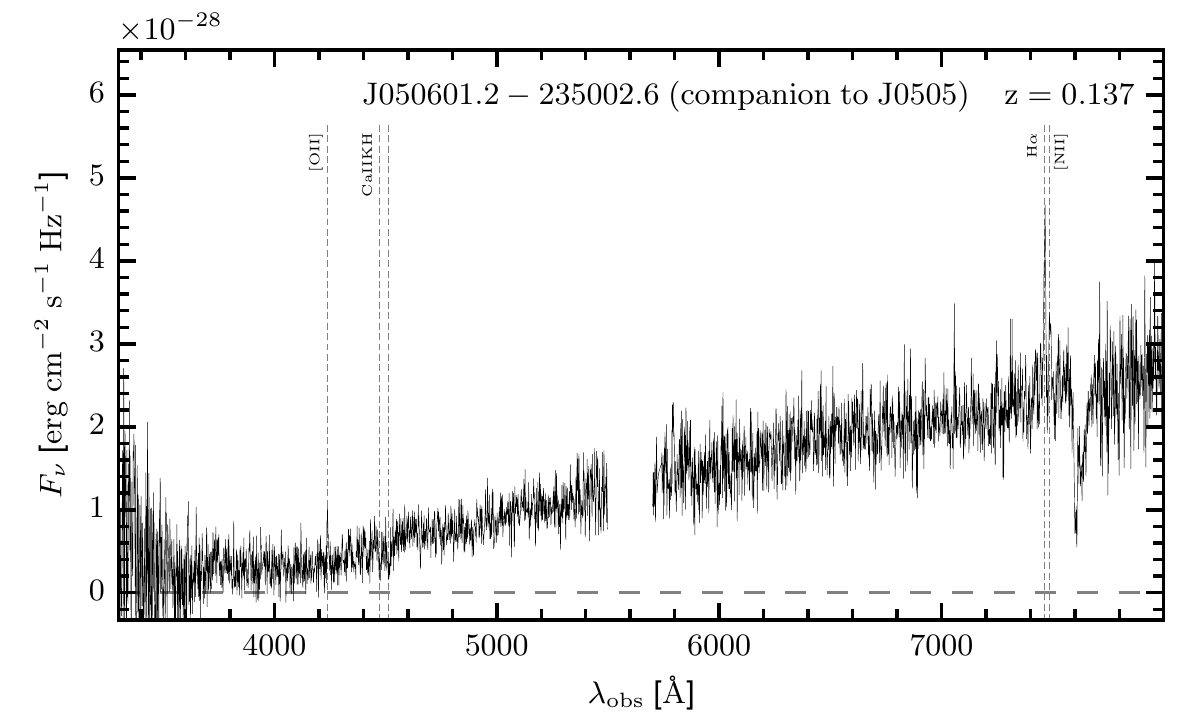}
\caption{Keck optical spectrum for J050601.2--235002.6, the apparent companion galaxy to
  2MFGC~04170 (the host galaxy for J0505). Multiple emission and absorption lines are
  identified, and labeled here.}
\label{opt_J0505comp} 
\end{figure}

As described in the main text, with the Keck telescope we performed
optical spectroscopy for J050601.2--235002.6, the apparent companion
galaxy to 2MFGC~04170 (the host galaxy for J0505). The resulting
spectrum is shown in Figure \ref{opt_J1506comp}. The relatively high
  redshift ($z=0.137$) confirms that this is a background galaxy and a chance
  alignment with 2MFGC~04170 ($z=0.036$).

\subsection{J1506}

\begin{figure}
\centering
\includegraphics[width=0.47\textwidth]{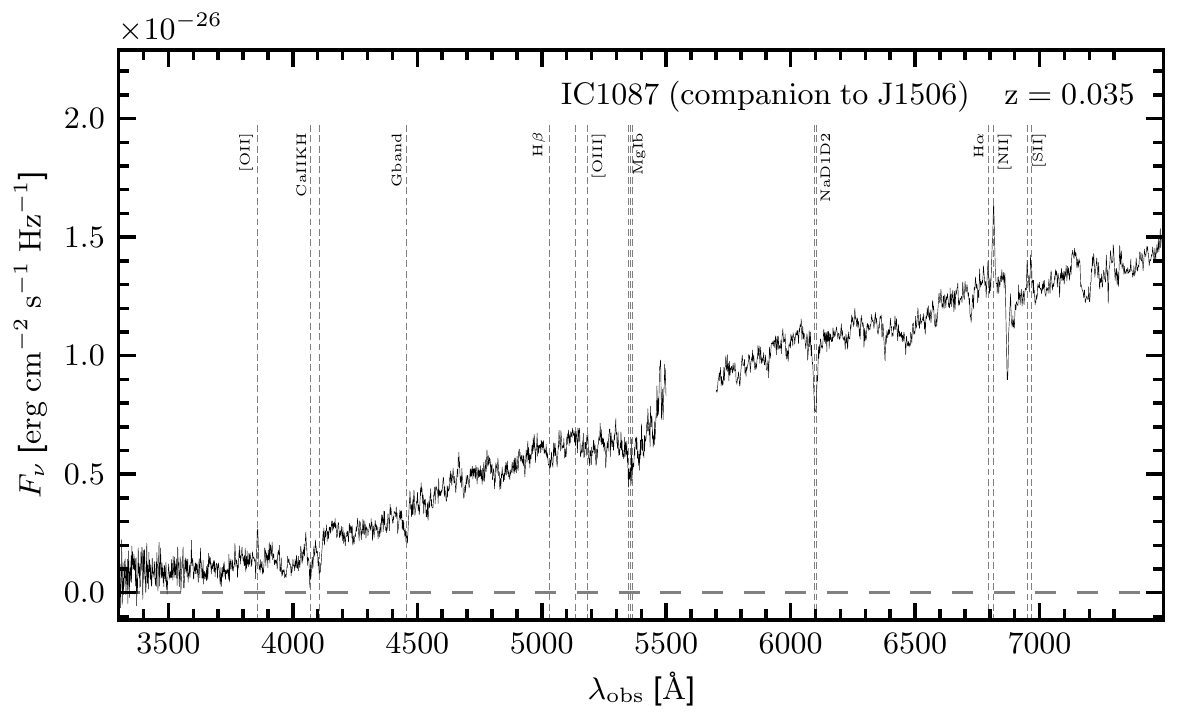}
\caption{Palomar optical spectrum for IC~1087, the merging companion galaxy to
  UGC~09710 (the host galaxy for our lowest redshift extreme \nustar
  source, J1506). Multiple emission and absorption lines are
  identified, and labeled here.}
\label{opt_J1506comp} 
\end{figure}

As described in the main text, J1506 belongs to one of two galaxies in
a major merger. With the Palomar observatory Hale telescope we
performed optical spectroscopy for the companion galaxy (known as
IC~1087). The resulting spectrum is shown in Figure \ref{opt_J1506comp}.

\subsection{J1534}

\begin{figure}
\centering
\includegraphics[width=0.47\textwidth]{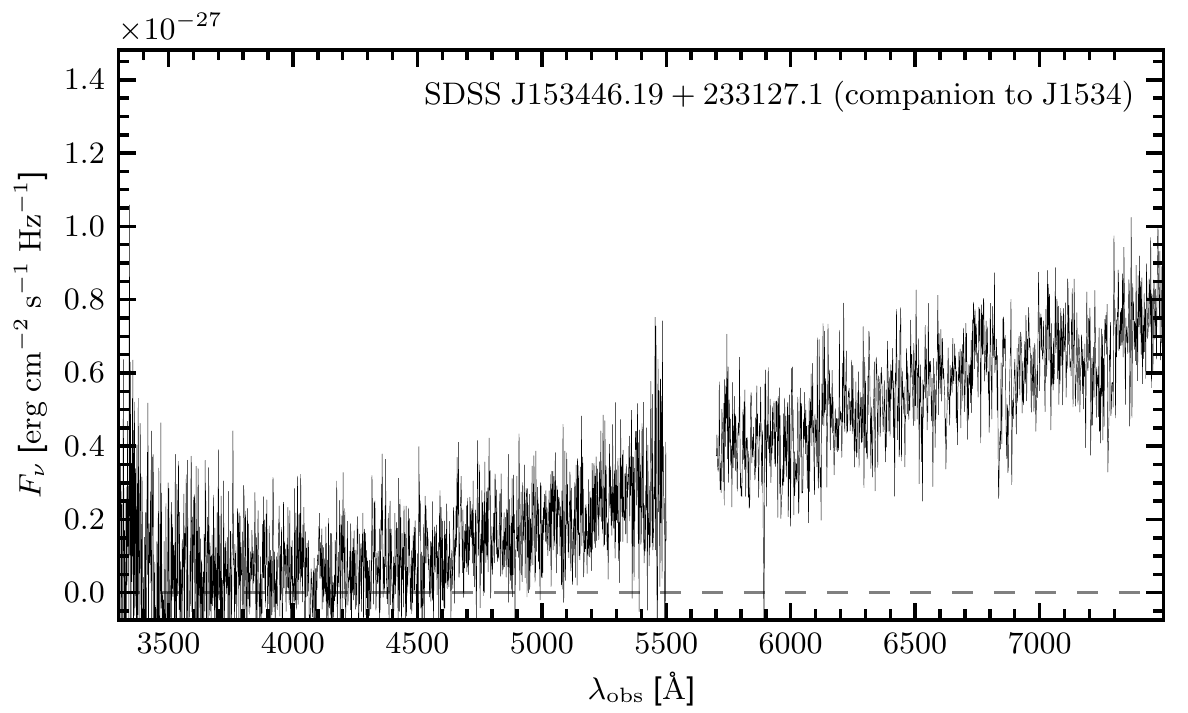}
\caption{Palomar optical spectrum for SDSS J153446.19+233127.1, the merging companion galaxy to
  SDSS J153445.80+233121.2 (the host galaxy for J1534). The continuum
  is detected, although no clear emission or absorption lines are
  identified, precluding a spectroscopic redshift measurement.}
\label{opt_J1534comp} 
\end{figure}

As described in the main text, J1534 (hosted by galaxy SDSS J153445.80+233121.2) appears to be undergoing a major
merger with a neighbouring galaxy (SDSS J153446.19+233127.1). Since no
spectroscopic redshift is available for the latter galaxy, we
performed optical spectroscopy with the Palomar observatory Hale
telescope, the spectrum from which is shown in Figure
\ref{opt_J1534comp}. Since no clear emission or absorption features
are detected, this companion requires deeper 
spectroscopic observations in the future to reliably determine the redshift.

%%%%%%%%%%%%%%%%%%%%%%%%%%%%%%%%%%%%%%%%%%%%%%%%%%%%%%%%%%%%%%%%%%%%%%

%%%%%%%%%%%%%%%%%%%%%%%%%%%%%%%%%%%%%%%%%%%%%%%%%%%%%%%%%%%%%%%%%%%%%%
\end{document}